\documentclass[a4paper,11pt]{article}
\pdfoutput=1
\usepackage{jheppub}
\usepackage{hyperref}
\usepackage{mathdots}
\usepackage{MnSymbol}
\usepackage{amsmath}
\usepackage{amsfonts}
\usepackage{graphicx}
\usepackage{caption}
\usepackage{subcaption}
\usepackage{placeins}
\usepackage{tikz}
\usepackage{svg}
\usepackage{subfiles}

\usepackage[english]{babel}

\usepackage{CJK}

\usepackage{array}
\usepackage{mathtools}
\def\be{\begin{equation}}
\def\ee{\end{equation}}
\def\ba{\begin{eqnarray}}
\def\ea{\end{eqnarray}}

\title{Tropical Amplitudes For Colored Lagrangians}

\author[a]{Nima Arkani-Hamed,}
\author[b]{Carolina Figueiredo,}
\author[c]{Hadleigh Frost,}
\author[d]{Giulio Salvatori}

\affiliation[a]{School of Natural Sciences, Institute for Advanced Study, Princeton, NJ, 08540, USA}
\affiliation[b]{Jadwin Hall, Princeton University, Princeton, NJ 08540, USA}
\affiliation[c]{Mathematical Institute, Andrew Wiles Building, Woodstock Rd, Oxford, UK}
\affiliation[d]{Max-Plank-Institüt fur Physik, Werner-Heisenberg-Institut, D-80805 München, Germany}

\emailAdd{arkani@ias.edu}
\emailAdd{cfigueiredo@princeton.edu}
\emailAdd{frost@maths.ox.ac.uk}
\emailAdd{giulios@mpp.mpg.de}

\abstract{Recently a new formulation for scattering amplitudes in Tr($\Phi^3$) theory has been given based on simple combinatorial ideas in the space of kinematic data. This allows all-loop integrated amplitudes to be expressed as ``curve integrals'' defined using tropical building blocks --- the ``headlight functions''. This paper shows how the formulation extends to the amplitudes of more general Lagrangians. We will present a number of different ways of introducing tropical ``numerator functions'' that allow us to describe general Lagrangian interactions. The simplest family of these ``tropical numerators'' computes the amplitudes of interesting Lagrangians with infinitely many interactions. We also describe methods for tropically formulating the amplitudes for general Lagrangians. One uses a variant of ``Wick contraction'' to glue together numerator factors for general interaction vertices. Another uses a natural characterization of polygons on surfaces to give a novel combinatorial description of all possible diagrams associated with arbitrary valence interactions.}

\begin{document}
\maketitle
\addtocontents{toc}{\protect\setcounter{tocdepth}{1}}
\section{Introduction}
Recent work has shown that the scattering amplitudes in the simplest theory of interacting colored scalars -- the Tr$(\Phi^3)$ theory -- can be expressed as ``curve integrals'' \cite{curveint}. This applies to all orders in the topological 't Hooft expansion. A given order in the expansion is associated with a surface, $\Sigma$, and every curve $C$ on this surface coincides with a propagator factor, $1/X_C$, that can arise in Feynman diagrams of the theory. To each such curve is associated a ``headlight function'', $\alpha_C({\bf t})$, which is a piecewise-linear function on an $E$-dimensional vector space with coordinates ${\bf t}$. Here, $E$ is the number of propagators in any Feynman diagram contributing to the order under consideration. Equivalently, $E$ is the dimension of the Teichm\"uller space associated to the surface $\Sigma$.\footnote{A surface with genus $g$ and $h$ boundaries contributes at order $L = 2g+h-1$ in the loop-order expansion. For $n$ external particles, $E = n-3 + 3L$. For an introduction to Teichm\"uller theory, see \cite{penner2012decorated}}

The amplitude ${\cal A}(\Sigma)$ associated to the surface $\Sigma$ can be expressed as a \emph{curve integral} over ${\bf t}$ space\footnote{The curve integral is a ``global Schwinger parametrization''. At loop level this allows the loop integrations to be performed leaving us with an integration over ${\bf t}$ space depending on ``surface Symanzik polynomials''. In addition, beginning with the annulus at one-loop, we have to further mod out by the action of the mapping class group of the surface, by including the ``Mirzakhani Kernel'' ${\cal K}(\alpha)$ as a factor in the integrand. These points are described in detail in \cite{curveint}; we suppress them in what follows to streamline our discussion.} 
\begin{equation}\label{eq:AS}
{\cal A}_{{\rm Tr}(\Phi^3)}(\Sigma) = \int {\rm d}^E {\bf t} \, {\rm exp} \left(- \sum_C X_C \alpha_C ({\bf t}) \right),
\end{equation}
where the sum is over all curves $C$.  We can integrate over the overall scale of the ${\bf t}$ variables to express the amplitude as an integral over an $(E-1)$-sphere as
\begin{equation}\label{eq:ASP}
{\cal A}_{{\rm Tr}(\Phi^3)}(\Sigma) =  \int_{S^{E-1}}  \frac{\langle t d^{E-1} t \rangle}{\left(\sum_C \alpha_C({\bf t }) X_C\right)^{E}},
\end{equation}
where $X=(P^2+m^2)$, the propagator factor\footnote{In the rest of the paper we often abuse notation and use $X$ to label both the curve as well as the propagator factor $X=P^2+m^2$.}.

In equations (\ref{eq:AS}--\ref{eq:ASP}), consider the ``action'', $S = -\sum_C X_C \alpha_C$. This sums over all possible curves, with no knowledge of which propagators are allowed to come together to form a Feynman diagram. The domains of linearity of $S$ correspond precisely to allowable Feynman diagrams, or, equivalently, to triangulations of the surface. This is remarkable because each $\alpha_X$ can be computed from knowing the curve $X$, without knowing the details of the rest of the surface or about other curves. Nevertheless, the linear domains of the action $S$ discover all the triangulations (or Feynman diagrams) of $\Sigma$.

Each linear domain of the action $S$ is a \emph{cone}. When restricted to such a cone, the integrals (\ref{eq:AS}--\ref{eq:ASP}) reproduce the Schwinger parametrizations of the associated Feynman diagram. The curve integral over the entire ${\bf t}$-space then produces the full amplitude. The cones in ${\bf t}$-space are spanned by ``$g$-vectors'', ${\bf g}_X$. Every cone is generated by the $E$ $g$-vectors of the curves defining the triangulation associated to the cone. The collection of all these cones defines a ``fan'' that covers all of ${\bf t}$ space. A lightning review of these topics is given in Appendix \ref{sec:review}. 

The formulas (\ref{eq:AS}--\ref{eq:ASP}) are for the scalar theory with interaction Lagrangian ${\cal L}_{\rm int} = {\rm Tr} (\Phi^3) = (\Phi^a_b \Phi^b_c\Phi^c_a)/3$. Our goal in this article is to show how the curve integral formalism extends to compute the amplitudes of theories with other Lagrangians. The generalization consists of adding a ``tropical numerator'' ${\cal N}$ to the curve integral
\begin{equation}\label{eq:ASPN}
{\cal A}_{{\cal L}}(\Sigma) =  \int_{S^{E-1}}  \frac{\langle t d^{E-1} t \rangle {\cal N_{\cal L}}}{\left(\sum_C \alpha_C({\bf t }) X_C\right)^{E}},
\end{equation}
which encodes the interactions of a Lagrangian ${\cal L}$\footnote{For other approaches to treat scalar fields with polynomial interactions in positive geometric fashion, see~\cite{jagadale2022towards,Kojima:2020tox,Srivastava:2020dly,Banerjee:2018tun,Aneesh:2019cvt,Jagadale:2023hjr,Kalyanapuram:2019nnf,Kalyanapuram_2020,Aneesh:2019ddi,Jagadale:2022rbl,Jagadale:2020qfa}.}.

We end our introductory remarks by giving a narrative road-map for the paper.

We begin in sections \ref{sec:general} and \ref{sec:tropnum} by describing the representation of amplitudes for general Lagrangians as a sum over cubic graphs, and the definition of ${\cal N}$ as a piecewise-constant function on the same fan we have in the curve integral for Tr$(\Phi^3)$ theory. We introduce new tropical functions that are close cousins of the headlight functions, the $\Theta_X$ variables, that the numerator functions ${\cal N}$ depend on. 

While the amplitudes for any Lagrangian can be trivially written in this way --- manually weighting each cubic diagram with its own numerator factor --- the goal is to find different presentations of the numerator function, without making any explicit reference to the Feynman diagrams/cones of the fan,  that achieve the same final result. This is both practically and parametrically important at large multiplicity $n$ when the number of cones/diagrams grows exponentially. In section \ref{sec:polexp} we emphasize that we are looking for a formulation of ${\cal N}$ that can be computed instead in polynomial time in $n$, echoing a well-known phenomenon in mathematics where analytic expressions with exponentially many terms can still be efficiently computed in polynomial time.\footnote{For several examples, see the first chapter of \cite{maclagan2021introduction}} In section \ref{sec:fact} we define the required properties of the numerator function ${\cal N}$($\Theta_X$,$X$) that ensure consistent factorization of the full integral. This will then guide the search for simple representations of ${\cal N}$ that will define physically sensible amplitudes. 

In section \ref{sec:generalCubic}, we give tropical formulations for completely general cubic Lagrangians in terms of the product of a simple function over all the triangles on the surface. While identifying all the triangles on a surface is intuitive and trivial at tree-level, we must learn how to systematically do it for a general surface. It turns out this can be achieved using the fat graph defining the surface. The method presented exposes a surprising bijection between the triangles of the base triangulation defining the surface and the triangles in any triangulation. We use this bijection to define ``tropical vertices'' which naturally provide a second tropical formulation for any cubic Lagrangian. 

Moving on to higher-valency Lagrangian interactions, in section \ref{sec:simplenonpoly} we discover that already at tree-level the simplest extension from products of triangles to more general products over $m$-gons does give consistently factorizing amplitudes, but for interesting Lagrangian with infinitely many interactions. This gives a different basis for the space of all possible Lagrangians than the conventional one associated with finite numbers of polynomial interactions. 

In section \ref{sec:mgons} we extend the definition of triangles to $m$-gons for a general surface and use this to define the amplitudes for these interesting Lagrangians at all loops. In section \ref{sec:partition}, we also show that this definition of $m-$gons allows for a striking combinatorial characterization of all possible polyangulations of the surface/Feynman diagrams with arbitrary valence interactions, in terms of partitions of the triangles in the base fat graph.  

Finally, in section \ref{sec:dictionary} we establish a straightforward dictionary between tropical numerators and general Lagrangians. We do this in two ways. In the first, we add further auxiliary variables $\phi_X$. Using the familiar idea of Wick contraction we build numerator functions that precisely produce the vertices for any desired Lagrangian ${\cal L}$. The formulas are very different from the conventional Wick contraction formulas that generate Feynman diagrams. Rather than generating the combinatorics of Feynman diagrams, Wick contractions on the $\phi_X$ variables compute numerator functions ``once for all diagrams''. The second method uses the striking combinatorial characterization of all Feynman diagrams in terms of partitions given in section \ref{sec:partition}, defining ``tropical vertices" for arbitrary valence interactions that automatically produce the correct numerators for the interaction vertices of any Lagrangian. The ``Wick'' method can also be used to extend our discussion to the amplitudes for multiple species of scalars/particles of higher spin. 

We end in section \ref{sec:outlook} with a discussion of some obvious open avenues for further exploration suggested by our results.

\section{General Scalar Theories and Cubic Numerators}
\label{sec:general}

The Feynman diagrams for Tr$(\Phi^3)$ theory correspond to triangulations $T$ of surfaces, with every diagram/triangulation contributing a factor $\prod_{X \in T} (1/X)$. For a given surface, $\Sigma$, we can compute the corresponding amplitude or integrand as a sum over these factors, for every distinct triangulation of $\Sigma$. In this section, we show that this description of amplitudes and integrands, --- as sums over triangulations, --- can be extended to the amplitudes of a general scalar Lagrangian. In subsequent sections, we show how this idea gives formulas for the amplitudes of such theories using curve integrals.

To begin, recall how Lagrangian interactions can be written in momentum space. Consider a single-trace Lagrangian interaction with $m$ fields of the form:
\begin{equation}
   \mathcal{L} = \text{Tr}(\Phi(k_1) \Phi(k_2) \cdots \Phi(k_m)) \times {\cal L}^{(m)}(k),
\end{equation}
for some function, ${\cal L}^{(m)}(k_1,\ldots,k_m)$, defined on the support of the momentum conservation relation: $\sum_a k_a^\mu = 0$. The cyclic symmetry of the trace tells us that the function ${\cal L}^{(m)}$ must be cyclically invariant. We can view the momenta $k^\mu_a$ as being associated with a momentum-polygon. The ${\cal L}^{(m)}$ are functions of invariant dot products $k_a \cdot k_b$, or, equivalently, they are functions of the variables
\begin{equation}
X_{a,b} = (k_a + k_{a+1} + \cdots k_{b-1})^2 + m^2,
\end{equation}
which as usual correspond to the length squares of the chords of the $m$-gon formed by the momenta $k_i$. 

In the rest of the paper and only for brevity we will set $m^2 = 0$, it is trivial to turn $m^2 \neq 0$ back on. Note that since the $k_a$ are off-shell, $X_{a,a+1} = k_a^2$ is not necessarily zero.  In other words, the $m$-valent interactions are represented by a cyclically invariant function ${\cal L}^{(m)}(X_{a,b})$ on the space of momentum $m$-gons.

For example, consider a cubic interaction defined by the function
\begin{equation}\label{eq:L3ex}
{\cal L}^{(3)} = (X_{a_1,a_2} + X_{a_2,a_3} + X_{a_3,a_1}),
\end{equation}
which, in position space, corresponds to the interaction vertex Tr$(\Phi \partial_\mu \Phi \partial^\mu \Phi)$. 

\subsection{Blowing up higher valence vertices}
\begin{figure}[t]
    \centering    
    \includegraphics[width=0.3\linewidth]{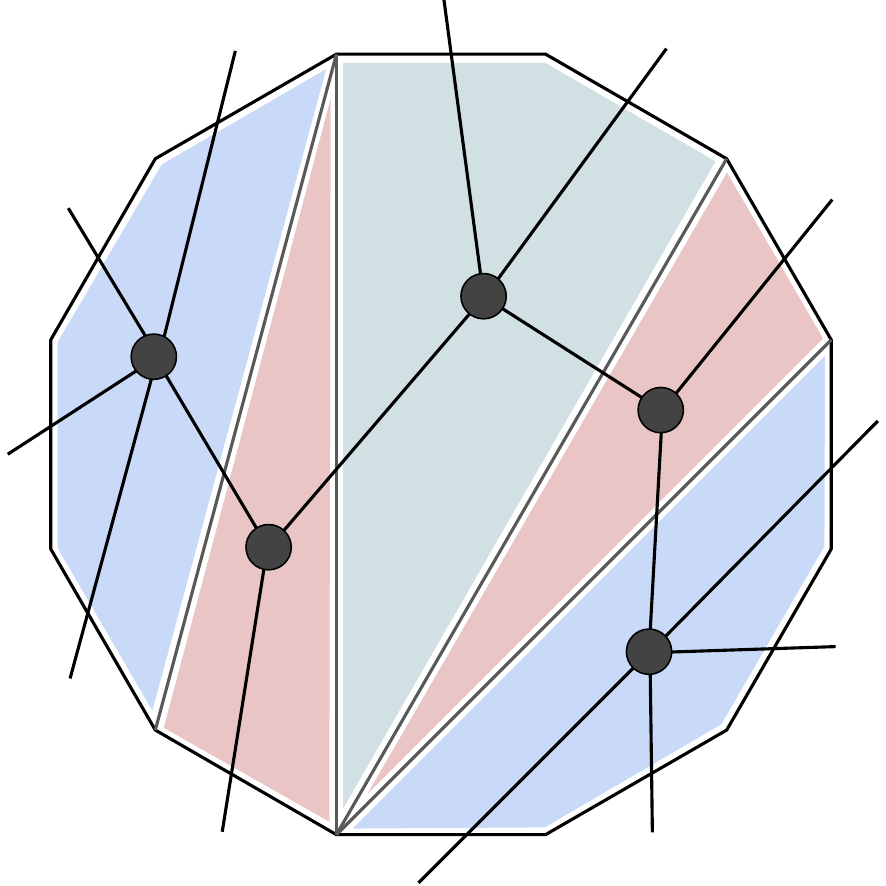}
    \caption{A ``polyangulation'' dual to a general Feynman diagram at tree level.}
    \label{fig:polylag}
\end{figure}

Consider now a single trace Lagrangian for a single colored scalar, ${\cal L}$, with some interaction terms, ${\cal L}^{(m)}$ ($m\geq 3$), as above. Any Feynman diagram for this Lagrangian is dual to a ``polyangulation'' of the surface with various $m$-gons, as in figure \ref{fig:polylag}. The Feynman rules assigned to this diagram give the product of $1/X$ for each internal chord of the polyangulation, multiplied by factors ${\cal L}^{(m)}$ for every $m$-gon in the polyangulation.

\begin{figure}[t]
    \centering    \includegraphics[width=\linewidth]{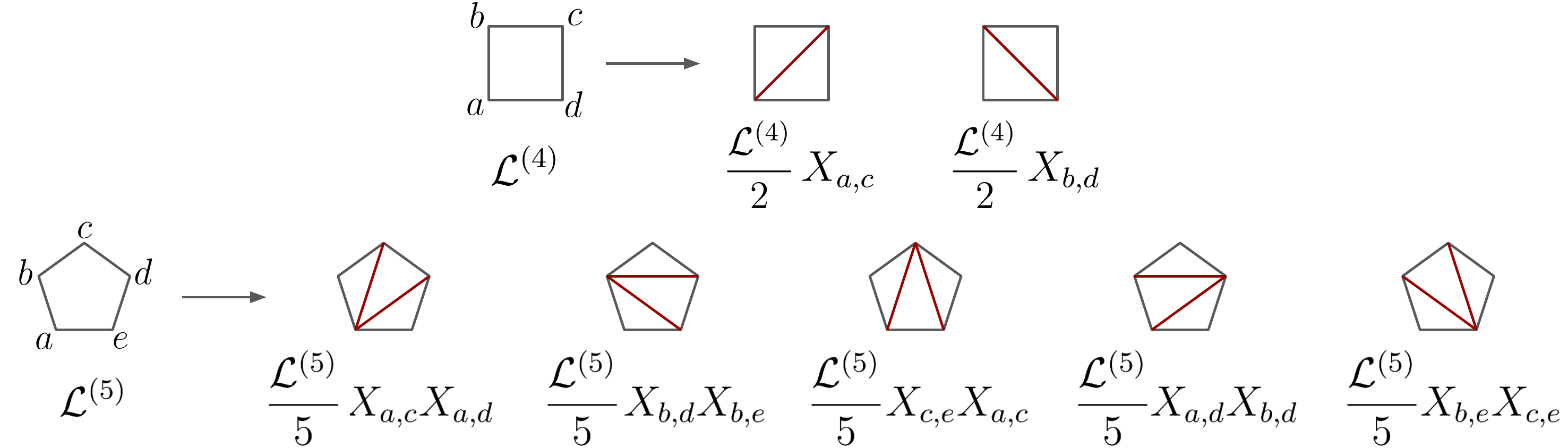}
    \caption{``Blowing up'' a four-point and five-point Lagrangian interaction ${\cal L}^{(4)}$ and ${\cal L}^{(5)}$, respectively, as an average over the two/five triangulations of the square/pentagon.}
    \label{fig:polygon}
\end{figure}

This sum over polyangulations can be rewritten as a sum over triangulations. This can be achieved by writing an $m$-valent interaction term ($m>3$) as a weighted sum over cubic diagrams. We can describe this replacement as ``blowing up'' the vertex. For instance, consider ${\rm Tr}(\Phi^4)$ amplitudes. The 4-point interaction can be rewritten as a sum: the $s$-channel cubic diagram multiplied by a factor of $s/2$, plus the $t$-channel diagram multiplied by a factor of $t/2$ (see figure \ref{fig:polygon}). Note that this ``blow up'' is not unique. Any such weighted average of these two channels will work. See figure \ref{fig:polygon} for another example of a blow up that writes a 5-point interaction as a sum of cubic interactions. When blowing up an $m$-point vertex, we can sum over triangulations $\tau$ of the $m$-gon with any weights, $w_\tau$, that satisfy $\sum_t w_\tau = 1$. To fix ideas, we could take, e.g., $w_\tau = 1/C_{m-2}$, where $C_{m-2}$ is the Catalan number that counts the number of triangulations of the $m$-gon.

Using this idea, we can write the Feynman rules for the general Lagrangian ${\cal L}$ as a sum over cubic diagrams. The final result will be a sum over triangulations, $T$,
\begin{equation}\label{eq:sumTNT}
\sum_T \frac{n_{{\cal L},T}}{\prod_{X\in T} X},
\end{equation}
for some numerators, $n_{{\cal L},T}$, and with the usual product of propagators in the denominator. The numerator, $n_{{\cal L},T}$, is given by summing over all polyangulations that are compatible with the chords of $T$. For each such polyangulation, we multiply together factors corresponding to each $m$-gon in the polyangulation. Each such $m$-gon is itself triangulated (according to $T$) by a smaller triangulation, $\tau$. For each of these $m$-gons, we include a factor of $w_\tau \, {\cal L}^{(m)}(X) \left(\prod_{X \in \tau} X\right)$, where ${\cal L}^{(m)}$ is a function of the $X$'s on the boundary of the $m$-gon, and the product is over $X$'s internal to the $m$-gon. With this definition of $n_{{\cal L},T}$, \eqref{eq:sumTNT} then reproduces the sum over all polyangulations with the original Feynman rules.

\subsection{Lagrangian Field Redefinitions}
We pause to comment on field redefinitions, which are transformations that change the Lagrangian while leaving the amplitude unchanged. In momentum space, this ambiguity is reflected in the fact that the ${\cal L}^{(m)}(X_{a,b})$ can depend on all $X_{a,b}$, including the $X_{a,a+1}$ which are zero on-shell. For instance, in the theory with just a cubic interaction given by \eqref{eq:L3ex}. The three-particle amplitude vanishes since $X_{1,2} = X_{2,3} = X_{3,1} = 0$, while the four-particle amplitude is ${\cal A}_4 = (X_{1,3} + X_{2,4})$. So, for this theory, we can make a field redefinition to set the cubic coupling to zero while generating quartic and higher terms to compensate.

Is there a canonical choice for how to present the Lagrangian? For colored theories in the language we are using here, there is a simple canonical Lagrangian.  Consider some $m$-particle tree-amplitude,
\begin{equation}
{\cal A}^{(m)}_{{\rm tree}} = {\cal A}^{(m)}_{{\rm poles}} + {{\cal A}}^{(m)}_{{\rm contact}},
\end{equation}
which we write as a sum of the terms with poles, ${\cal A}^{(m)}_{{\rm poles}}$, and the ``contact terms'', ${{\cal A}}^{(m)}_{{\rm contact}}$, which have no poles. Note that these are functions only of the internal chords of the $m$-gon, since the boundary $X_{a,a+1}$ are zero on-shell. We can then use this function of the internal $X$'s of the $m$-gon to define a canonical Lagrangian, ${\cal L}_{{\rm amp}}$, determined directly by the on-shell amplitudes via ${\cal L}_{{\rm amp}}^{(m)} = {\cal A}^{(m)}_{{\rm contact}}$.

For example, in the simple cubic theory above, with cubic vertex \eqref{eq:L3ex}, there are non-zero contact terms in all the tree amplitudes for all $n\geq 4$. The canonical Lagrangian for this theory is found by computing these contact terms, and it has vertices (written in momentum space) 
\begin{equation}
    {\cal L}^{(m)}_{{\rm amp}} = \sum_{i,j} X_{i,j} C_{|i-j| - 2} C_{|m - |i-j|| - 2},
\end{equation}
for every $m\geq 3$, with $C_{-1},C_0 = 1$.

\section{Tropical Numerators and $\Theta$ functions}
\label{sec:tropnum}

The sum, \eqref{eq:sumTNT}, which reproduces the Feynman rules of a Lagrangian ${\cal L}$, can also be written as a curve integral,
\begin{equation}\label{eq:ALNL}
{\cal A}_{\cal L} =  \int\limits_{S^{E-1}}  \frac{\langle t d^{E-1} t \rangle \, {\cal N}_{{\cal L}}({\bf t})}{(\sum_X \alpha_X({\bf t}) X)^{E}},
\end{equation}
where we have introduced a \emph{tropical numerator function} ${\cal N}_{{\cal L}}({\bf t})$. This is a {\bf piecewise constant} function on ${\bf t}$-space. In a cone, $C$, associated to a triangulation of the surface, we define ${\cal N}_{{\cal L}}$ to take the value $n_{{\cal L},C}$. To write ${\cal N}_{\cal L}$ explicitly, introduce tropical $\Theta$-functions, $\Theta_C$, for each cone $C$, that satisfy
\begin{equation}
\Theta_C({\bf t}) = \begin{cases} 1 \quad {\rm if } \, {\bf t} \, \in C, \\ 0 \quad {\rm otherwise.} \end{cases}
\end{equation}
Then the tropical numerator function can be written as
\begin{equation}\label{eq:NL}
{\cal N}_{{\cal L}}({\bf t}) = \sum_C n_{{\cal L},C}\, \Theta_C({\bf t}),
\end{equation}
where we sum over all cones, $C$. This formula for $\mathcal{N}_\mathcal{L}$ is no better than summing over all Feynman diagrams. However, in subsequent sections, we will show how to efficiently compute tropical numerator functions, \emph{without} summing over all cones.

\subsection{Tropical $\Theta$ Functions}

The functions, $\Theta_C$, can be efficiently computed using the headlight functions, $\alpha_X$. To find a formula for the $\Theta_C$'s, consider first defining some new $\Theta$-functions $\Theta_X$, for each {\bf curve} $X$, which satisfy
\begin{equation}
\Theta_X = \begin{cases} 1  \quad {\rm in \, any \,  cone \, that \, contains} \,  X, \\ 0 \quad  {\rm otherwise.} \end{cases}
\end{equation}

Given the $\Theta_X$ functions, we can write the function $\Theta_C$, for any cone $C$, as
\begin{equation}
\Theta_C = \prod_{X \subset C} \Theta_{X},
\end{equation}
where the product is over every curve $X$ that generates the cones $C$.

We can find formulas for $\Theta_X$ using the headlight functions. Recall that the headlight function $\alpha_X$ is a piecewise-linear function that ``lights up'' the ${\bf g}$-vector for the curve $X$: $\alpha_X({\bf g}_Y)=\delta_{X,Y}$. Consider a cone $C$ spanned by the variables $X_1, \cdots, X_E$. In this cone, $\alpha_X$ vanishes identically unless $X$ is one of $X_1,\ldots,X_E$. For $\alpha_{X_a}$ we find (since the $\alpha_X$'s are linear) that
\begin{equation}
\alpha_{X_a} = t_i ({\bf G}^{-1})^{i}_a ,
\label{equation:dumbalpha}
\end{equation}
where the matrix ${\bf G}$ is formed from collecting all the $g$-vectors of the cone in column vectors as ${\bf G}^a_i = g_{{X_a} \, i}$. Since $\alpha_X$ is piecewise linear on every cone, the gradient $\partial_i \,\alpha_X$  is piecewise constant. It follows that ${\bf g}_X \cdot \nabla \alpha_X = 1$ in the cones that contain $X$, and $=0$ otherwise. Hence
\begin{equation}
\Theta_X = {\bf g}_X \cdot \nabla \alpha_X.
\label{equation:dumbtheta}
\end{equation}

As explained in the next section, both ${\bf g}_X$ and $\alpha_X$ can be computed efficiently using the curve integral formalism so that we can obtain simple formulas for $\Theta_X$ without having to consider evaluating it cone-by-cone (as in, e.g., \eqref{equation:dumbalpha}).

Finally, it will be useful to note that we can define an $E\times E$ matrix $\omega_X$, for every curve $X$, by
\begin{equation}
(\omega_X)^i_j =\frac{\partial \alpha_X}{\partial t_i} \, g_{X \, j},
\end{equation}
such that $\Theta_X$ is the trace: $\Theta_X = (\omega_X)^i_i$.
The $\omega_X$ are non-vanishing only in the cones containing $X$. It is also easy to see, using the properties of the headlight functions, $\alpha_X$, that the $\omega_X$ form a set of projectors:
\begin{equation}
\omega_X^2 = \omega_X, \qquad \sum_X \omega_X = 1, \qquad \omega_X \omega_Y = 0, \quad \text{for }X\neq Y.
\end{equation}

\begin{figure}[t]
    \centering
    \includegraphics[width=0.4\textwidth]{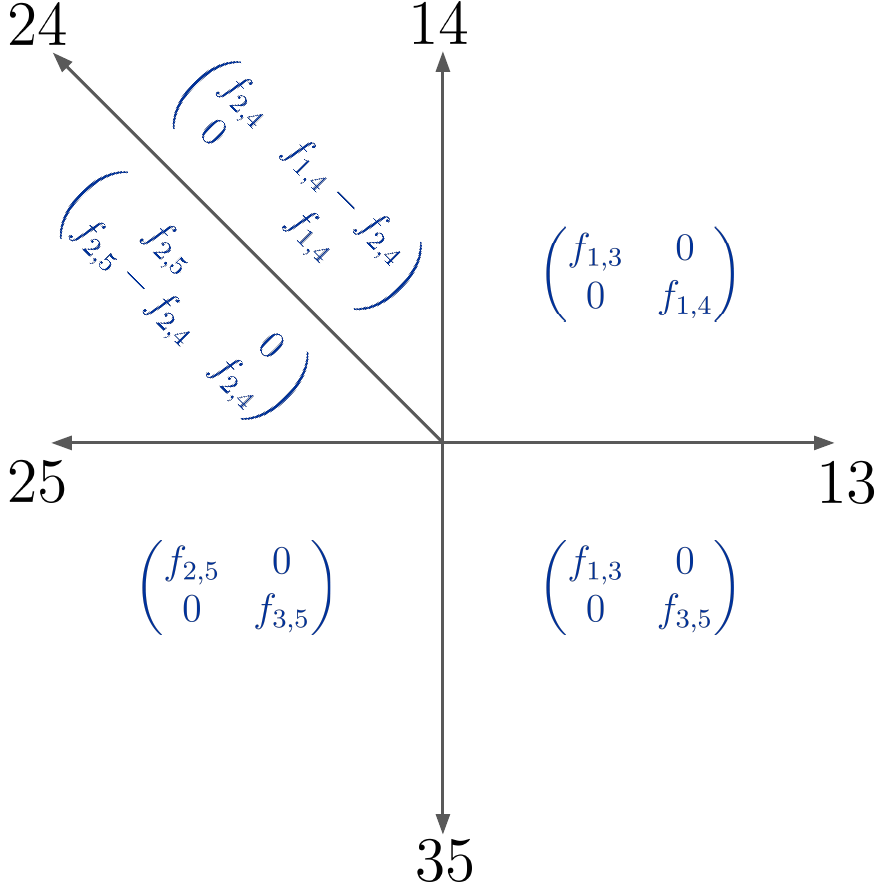}
    \caption{The weighted sum $\Omega = \sum_{i,j} f_{X_{i,j}} \omega_{X_{i,j}}$ of the projector matrices $\omega_{X_{i,j}}$, for the $n=5$, tree-level fan.}
    \label{fig:omega}
\end{figure}

We can collect all the $\omega_X$ into a single matrix $\Omega$ by introducing a dummy variable $f_X$ for every curve $X$, and define $\Omega = \sum_X f_X \omega_X$.
We show $\Omega$ for the simple case of the $n=5$ tree-level fan in figure \ref{fig:omega}. We will see a nice interpretation for det$(\Omega)$ in a moment. 

\subsection{Efficient tropical computation of $\alpha_X$ and $\Theta_X$}
The definitions of $\alpha_X$ and $\Theta_X$, via (\ref{equation:dumbalpha}) and (\ref{equation:dumbtheta}), could have been given for a random collection of rays and cones filling up ${\bf t}$ space. However, defining $\alpha_X$ in this way requires us to first enumerate cones, since \eqref{equation:dumbalpha} is given cone-by-cone. Using these definitions (essentially tautologically) allows us to write general amplitudes as an integral over ${\bf t}$ space, but they are in no way better than enumerating Feynman diagrams, since determining $\alpha_X, \Theta_X$ this way is as hard as finding all the cones/diagrams to begin with. 

However, there is an extra magic associated with our special setting of curves of surfaces \cite{curveint}. Instead of the direct and useless equation (\ref{equation:dumbalpha}), there are simple formulas for the $g$-vectors, ${\bf g}_X$, and headlight functions, $\alpha_X$, that are directly computed from the data specifying the curve $X$, with no reference to the cones that $X$ belongs to (see Appendix \ref{sec:review}). A further striking feature is that, when computing $\alpha_X$, the dependence on the number of external particles $n$ and the loop order, $L$, can be effectively decoupled. This means that computations for the simplest ``tadpole-like'' surfaces at any loop order can be used to immediately compute the $\alpha_X$ at all multiplicities as in \cite{alln}. 

Since both $\alpha_X$ and ${\bf g}_X$ can be efficiently computed from the data of $X$, the same is true of the functions $\Theta_X={\bf g}_X \cdot \nabla \alpha_X$. In Appendix A we also give an alternative tropical formula for $\Theta_X$.

\section{Polynomial vs. exponential complexity}
\label{sec:polexp}

Earlier we presented the amplitude for a Lagrangian ${\cal L}$ as a curve integral, \eqref{eq:ALNL}, and gave a formula for the tropical numerator function, ${\cal N}_{\cal L}$ as a sum over triangulations, \eqref{eq:NL}. By writing ${\cal N}_{\cal L}$ as a diagram-by-diagram sum, we are wasting the magic of the curve integral formulation. In the rest of this paper, we study ways to compute ${\cal N}_{\cal L}$ without this sum over diagrams. We are especially interested in formulas for ${\cal N}_{\cal L}$ that can be computed in polynomial time, but whose expansion (as a sum of terms) reproduces the exponentially-growing (with $n$) number of diagrams contributing to any given amplitude.

This is a general phenomenon. An example is the determinant of an $n \times n$ matrix, whose symbolic expression has $n!$ terms, but which can be computed in a time of order $n^3$ by Gaussian elimination. Another class of examples is given by products of a large number of small polynomials, such as 
\begin{equation}
    P(t,x)=(1 + t x_1) (1 + t x_2) \cdots (1 + t x_n).
\end{equation}

The symbolic expansion of this $n$'th degree polynomial in $t$ has $2^n$ terms. But, for any fixed numerical values of the $x_i$, $P(t,x)$ can be computed by iteratively multiplying the factors in a time of order $n^2$. We will see analogs of both of these examples in the following sections, where we study efficient formulas for ${\cal N}_{\cal L}$, expressed in terms of the $\Theta_X$ functions. 

\subsection{Cone-generating Functions as Determinants}
We illustrate an easy example of this polynomial vs. exponential phenomenon. Consider a ``cone generating function'', $F$, defined by the following sum over all cones:
\begin{equation}
F = \sum_{\substack{\text{cones}\\ \{X_{i_1}, \ldots, X_{i_E}\}}} f_{X_{i_1}} \cdots f_{X_{i_E}} \Theta_{X_{i_1}} \cdots \Theta_{X_{i_E}},
\end{equation}
for some variables, $f_X$, associated to each propagator, $X$. The number of terms in this sum grows very large with the number of particles, $n$. Even at tree level, there are $\sim 4^n$ (i.e. exponentially many) terms in the sum defining $F$. Despite this, there is a fast way to compute $F$ by expressing it as a determinant, which can be computed using Gaussian elimination. We find that $F = \det \Omega$, where the $E\times E$ matrix $\Omega$ defined above
\begin{equation}
\Omega^i_j = \sum_X f_X (\omega_X)^i_j,
\end{equation}
over all propagators, $X$. To see this result, recall that the $\omega_X$ matrices define a complete set of orthogonal projectors in every cone. It follows that, in a cone $\{X_1,\ldots,X_E\}$, we have $\det \Omega = f_{X_1} \cdots f_{X_E}$, which is the value of the function $F$ in that cone.

The number of curves, $X$, grows much more slowly than the number of diagrams. At tree level, the number of curves grows as $\sim n^2$, whereas the number of diagrams grows as $\sim 4^n$. It follows that the number of terms in the sum defining $\Omega$ grows slower than the number of terms in the definition of $F$. Moreover, at tree level, $E=n-3$, and $\Omega$ is an $E\times E$ matrix. It follows that $F$ can be computed using the formula $\det \Omega$ in polynomial time in $n$.

As an application of this result, note that the cone-generating function gives a new way to express amplitudes as curve integrals. If we set $f_X = 1/X$, then $F$ reproduces the product over propagators in each cone. It follows that the Tr $\Phi^3$ amplitude may be written as 
\begin{equation}
{\cal A}_{\text{Tr}(\Phi^3)}= \int_{S_{E-1}} \frac{\langle t d^{E-1} t \rangle \, {\rm det}\left(\sum_X \frac{1}{X} \Theta_X \right)}{\left(\sum_X \alpha_X\right)^E}.
\end{equation}
This expression has certain advantages to  \eqref{eq:ASP}; for instance, for many families of amplitudes the function in the denominator, $\sum_X \alpha_X$, has telescopic cancellations and simplifies dramatically. On the other hand, the $X$ dependence is not manifestly in Schwinger-parameteric form so loop integration is not trivialized as in \eqref{eq:ASP}. It is interesting to see that even in the simplest case of Tr$(\Phi^3)$ theory, we can have different tropical presentations of precisely the same amplitude. We will be seeing further examples of this in later sections for more general amplitudes. 

\section{Factorization}
\label{sec:fact}

We now turn to the problem of finding tropical numerator functions, ${\cal N}$, without summing over individual numerator factors for the different diagrams as in \eqref{eq:NL}. In general, we will consider functions ${\cal N}$ that are built out of the propagators $X$ and the functions $\Theta_X$. (Although in \cite{tropnlsm} we show that, for special theories, ${\cal N}$ can also be expressed directly as functions of the headlight functions, $\alpha_X$.)

In order for a function ${\cal N}$ to consistently define amplitudes, the main constraint is that ${\cal N}$ should respect factorization. If we regard ${\cal N}$ as a function ${\cal N}(X;\Theta)$ of $X$'s and $\Theta$'s, correct factorization on a pole, $X$, implies that
\begin{equation}
\lim_{X\rightarrow 0}\lim_{\substack{\Theta_X \rightarrow 1\\ \Theta_Y \rightarrow 0}} {\cal N}(X; \Theta) = {\cal N}^{{\rm cut}}(X;\Theta),
\nonumber
\end{equation}
where we send $\Theta_Y \rightarrow 0$ for every curve $Y$ that intersects $X$. In this formula, ${\cal N}^{{\rm cut}}$ is the numerator function for the surface (or surfaces) obtained by cutting along $X$. For example, at tree level, the amplitude factorizes on every pole into a ``left'' and ``right'' part. So the tropical numerators must satisfy \begin{equation}
{\cal N}^{{\rm cut}} = {\cal N}_{{\rm Left}}(\Theta_{{\rm Left}}) \times {\cal N}_{{\rm Right}}(\Theta_{{\rm Right}}).
\end{equation}

Simply by imposing that ${\cal N}$ factorizes correctly, we can produce amplitudes for a large class of theories. In particular, we will see that the amplitudes for \emph{every} single-trace scalar Lagrangian, ${\cal L}$, are recovered in this way.

\section{General cubic Lagrangians}
\label{sec:generalCubic}

Before discussing general Lagrangians, consider the case of cubic Lagrangians with a general interaction vertex, ${\cal L}^{(3)}$. In this section, we present two different tropical representations for the amplitudes of these theories. These formulas already contain the key ideas necessary for the general case.

\subsection{Tree Level}
For simplicity, we first explain how to compute tropical numerators ${\cal N}$ at tree level, and study the generalization to all loop orders separately. At tree level, every triangle that can arise in a triangulation, $C$, can be specified by its triple of vertices, $(a,b,c)$, and its associated propagators are $X_{a,b}, X_{b,c},X_{c,a}$.

Consider some (cyclically invariant) function, $g^{(3)}(X_{a,b},X_{b,c},X_{c,a})$, for every triangle. Then we define a tropical numerator function
\begin{equation}\label{eq:treeN3}
{\cal N}^{(3)} = \prod_{(a,b,c)} \left[1 + g^{(3)}(X_{ab},X_{bc},X_{ca}) \Theta_{X_{a,b}} \Theta_{X_{b,c}} \Theta_{X_{c,a}} \right],
\nonumber
\end{equation}
where the product is over all triangles $(a,b,c)$.  In this expression, let $\Theta_{X_{a, a+1}} =1$, for every boundary curve, $X_{a,a+1}$.\footnote{This is consistent with the defining property of the $\Theta$ functions, since boundary curves are compatible with every other curve.}

The function ${\cal N}^{(3)}$ factorizes correctly and therefore defines amplitudes. To see that it factorizes, set $\Theta_X \to 1$ for some $X$, and $\Theta_Y \to 0$ for all $Y$ that cross $X$. Then factor for any triangle $(a,b,c)$ that crosses $X$ becomes $1$. It follows that, in this limit, ${\cal N}^{(3)}$ becomes a product over only those triangles that do not intersect $X$. Since $X$ cuts the disk into two disks, in this limit ${\cal N}^{(3)}$ becomes
\begin{equation}
{\cal N}^{(3)} \rightarrow {\cal N}^{(3)}_{\rm left}\times {\cal N}^{(3)}_{\rm right},
\end{equation}
where ${\cal N}^{(3)}_{\rm left}$ is the product over all triangles on one side of $X$, and ${\cal N}^{(3)}_{\rm right}$ is the product over all triangles on the other side.

Finally, the amplitudes defined by ${\cal N}^{(3)}$ can trivially be understood as coming from the Lagrangian ${\cal L}^{(3)} = ( 1 + g^{(3)}(X))$.  In any cone corresponding to a triangulation, $C$, most of the factors in the product defining ${\cal N}^{(3)}$ are $1$. The only triangles that are $\neq 1$ are the triangles of $C$. Moreover, the factor for each of these triangles is $(1 + g^{(3)}(X))$, reproducing the correct numerator factor for this Lagrangian.

Note that the product expression for ${\cal N}^{(3)}$ is considerably simpler to evaluate than the sum over the exponentially many Feynman diagrams in \eqref{eq:NL}. Indeed, the number of terms in this product is the number of triangles, which grows only polynomially in $n$ ($\sim n^3$). To evaluate ${\cal N}^{(3)}$ we take a product over all triangles with \emph{no} reference to how they fit together into triangulations.
\begin{figure}[t]
    \centering
    \includegraphics[width=0.5\textwidth]{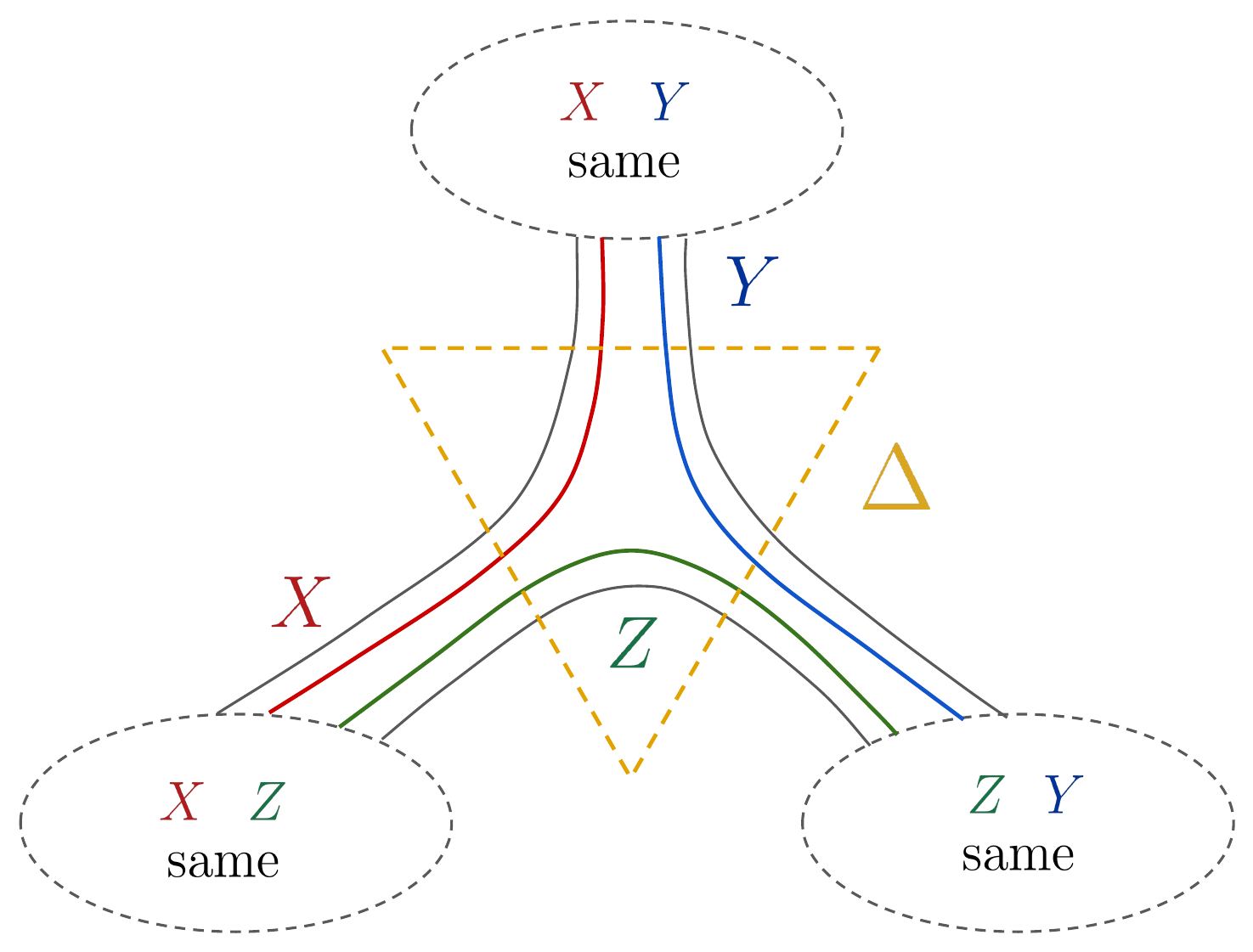}
    \caption{A cyclically ordered triple of curves $(X,Y,Z)$ on a surface defines a triangle if they meet on a vertex of the fat graph and look identical in pairs away from the vertex.}
    \label{fig:tri}
\end{figure}

\subsection{Triangles on a general surface}\label{sec:cubic:triangles}
The numerator function above, equation \eqref{eq:treeN3}, uses the fact that, at tree level, triangles are uniquely specified as cyclic tuples of points $(a_1,a_2,a_3)$. To find numerator functions valid to all loop orders, we have to first pause to understand how to describe a ``triangle'' for any surface in the topological expansion. How can we recognize a ``triangle'' on a general surface, and how can we generate them all systematically? 

The arena for all the action in the curve-integral formalism is the choice of a distinguished fatgraph that specifies (by duality) a particular triangulation of the surface. 
We'll now see how to use this distinguished fatgraph to define what we mean by a triangle on a general surface, and to systematically generate these triangles.

\begin{figure}[t]
    \centering
    \includegraphics[width=\textwidth]{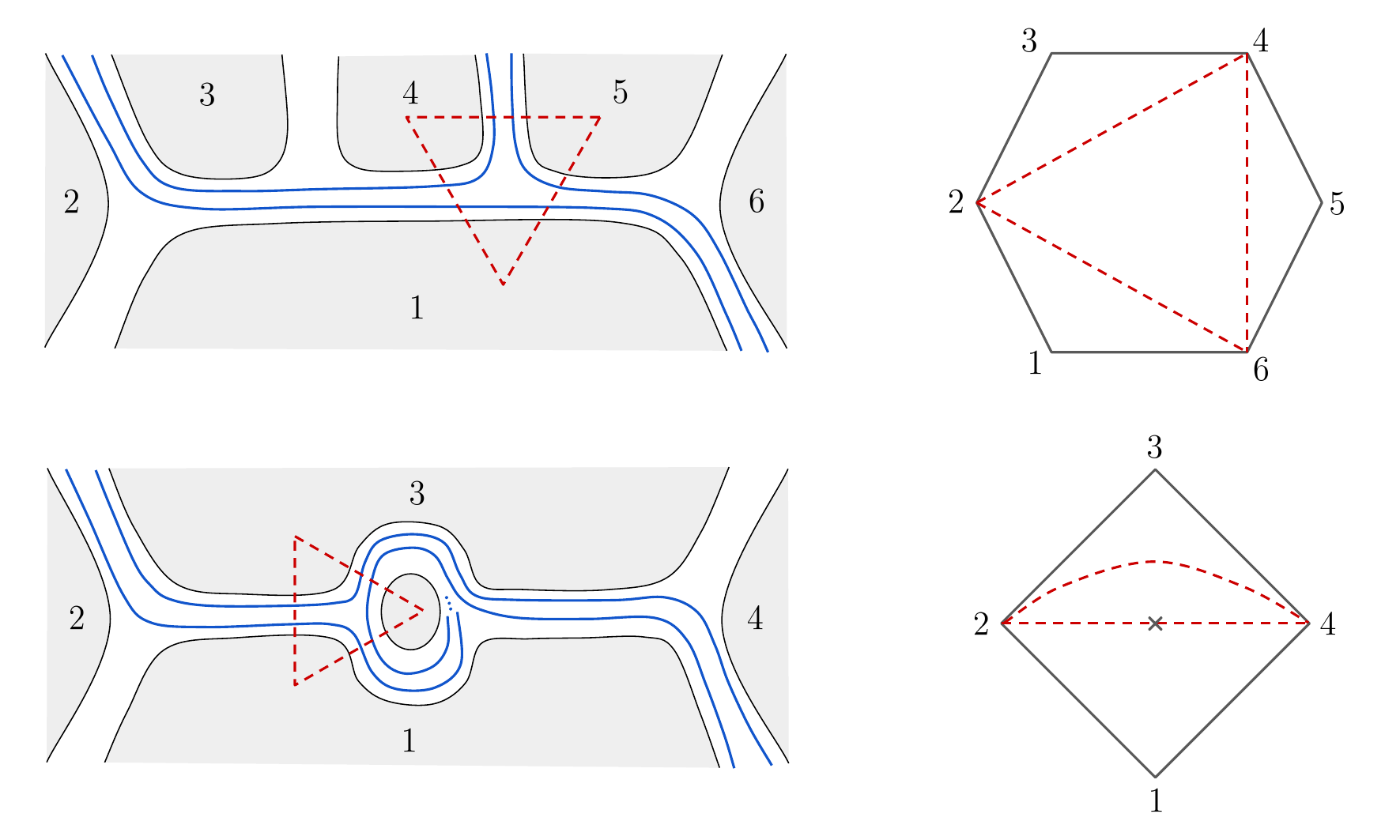}
    \caption{Identifying two triangles, at tree-level and one loop. The latter shows that spiraling curves are permitted to re-enter the triangle infinitely often. }
    \label{fig:tritreeloop}
\end{figure}

A triangle is defined by a cyclic triple of non-intersecting curves, $(X,Y,Z)$. But not every such triple defines a triangle. A triple defines a triangle if cutting the fatgraph along the curves cuts out a (fat) 3-vertex. This condition can be translated into a simple rule. There must be one vertex of the fatgraph where the 3 curves (respectively) hug the 3 sides of that vertex; and away from that vertex, the curves must run parallel to each other. This is shown in figure \ref{fig:tri}. Two concrete examples at tree-level and one-loop are shown in figure \ref{fig:tritreeloop}. 

To be more precise, we can express this rule in terms of the paths taken by the curves. For a triple $(X,Y,Z)$ to form a triangle, there must be a vertex of the fat graph, bounded by some roads $(i,j,k)$, such that
\begin{equation}\label{eq:XYZ}
X=A\,i\,L\, j\,\bar{B},\quad Y=B\,j\,L\,k\,\bar{C},\quad Z = C\,k\,L\,i\,\bar{A}, 
\end{equation}
for some paths $A,B,C$. Here $\bar{A},\bar{B},\bar{C}$ denote the reversed paths, and $L$ denotes a left turn. In fact, we can reverse engineer this definition of triangles to \emph{construct} all the triangles on a surface. We do this by starting from each vertex $(i,j,k)$ of the fat graph, and then building curves $(X,Y,Z)$ given by words of the above form, as in figure \ref{fig:tri}. 

\begin{figure}[t]
    \centering
    \includegraphics[width=0.8\textwidth]{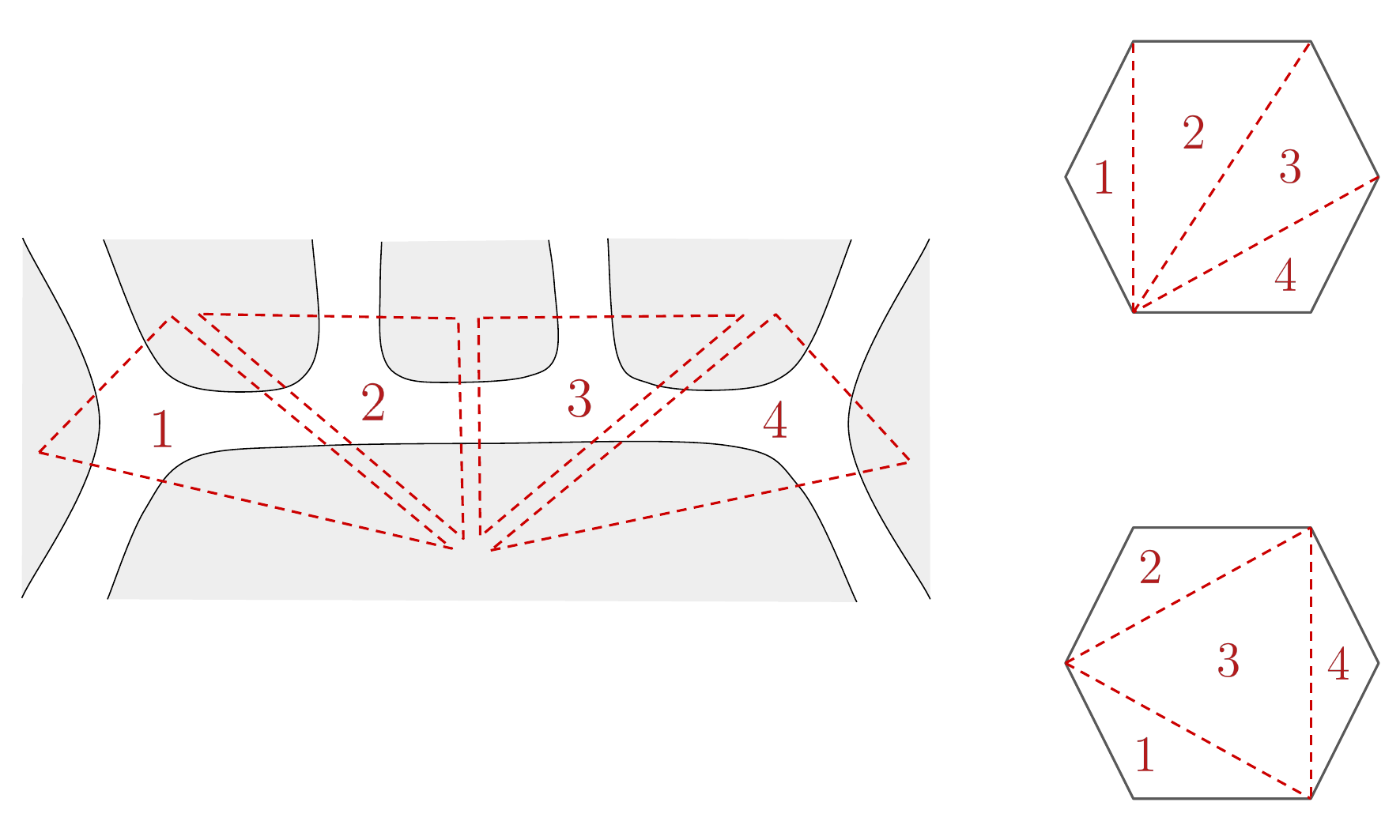}
    \caption{A base triangulation of the six particle tree, and the bijection between its triangles and those of a different triangulation. Note that adjacent triangles in the base need not have adjacent images.}
    \label{fig:bijection}
\end{figure}

This means that any given vertex of the fat graph --- or, dually, given a triangle $\Delta$ of the base triangulation $T$ --- is associated with a family of triangles on the surface. Any two distinct triangles in the family associated with some $\Delta$ must intersect each other.\footnote{If they did not intersect, then the paths $A,B,C$ associated with each triangle would have to be identical.} This means that, for any triangulation $C$ (associated to some cone), there is a 1:1 correspondence between the triangles $(X,Y,Z)$ of $C$ and the triangles $\Delta$ of the base triangulation $T$. We illustrate this with an example at tree-level in figure \ref{fig:bijection}. We will write $(X,Y,Z)_\Delta$ to denote a triangle $(X,Y,Z)$ that corresponds to (i.e. is dual to) the triangle $\Delta$, in the sense of \eqref{eq:XYZ}.

As illustrated in this example, the bijection is non-trivial: if two triangles $\Delta_1, \Delta_2$ in $T$ meet along an edge, this will not necessarily be the case for their images $\Delta_1^\prime, \Delta_2^\prime$ in $T^\prime$.

\subsection{All Loop Orders} 
Now that we know how to define triangles for general surfaces, we can define numerator functions for general cubic theories at all orders in the perturbation series. As in \eqref{eq:treeN3}, we can define ${\cal N}^{(3)}$ as a product over all triangles, which can now be written as
\begin{equation}\label{eq:loopN3}
{\cal N}^{(3)} = \prod_{\Delta \in T } \prod_{(X,Y,Z)_\Delta} \left[1 + g^{(3)}(X,Y,Z) \Theta_{X} \Theta_{Y} \Theta_{Z} \right],
\nonumber
\end{equation}
where the product is over all triples $(X,Y,Z)_{\Delta}$ (with the form \eqref{eq:XYZ}) that define triangles dual to each of the triangles $\Delta$ in $T$. At $L$ loops, there are $n-2+2L$ triangles $\Delta$ in the base triangulation. In a given cone, $C$, most of the factors in this product are $1$; there are only $n-2+2L$ factors $\neq 1$, which correspond to the $n-2+2L$ triangles $(X,Y,Z)_\Delta$ (for distinct $\Delta$) of the triangulation $C$.\footnote{At higher loop-level, there are in principle infinitely many factors in the product defining ${\cal N}^{(3)}$. But in practice, only $\sim (n+2L)^3$ factors correspond to triangles formed out of curves that are compatible with a choice of Mirzakhani kernel. So in computations ${\cal N}^{(3)}$ will always be a finite product. For the same reason, only a finite number of curves ever contribute to the curve integral for Tr$(\Phi^3)$ theory, as explained in \cite{curveint}.}

\subsection{Tropical Vertices}\label{sec:cubic:vertices}
The bijection between the triangles in the base triangulation, $T$, and triangles in any other triangulation, $C$, is an elementary but rather surprising and deep fact. This fact suggests an alternative way to define the tropical numerator functions.

Instead of the infinite product in equation \eqref{eq:loopN3}, consider defining a tropical numerator function as a finite product,
\begin{equation}\label{eq:loopN3V}
{\cal \tilde N}^{(3)} = \prod_{\Delta \in T} {\cal V}^{(3)}_\Delta,
\end{equation}
over the $n-2+2L$ triangles $\Delta$ in the base triangulation, $T$, where we introduce a ``tropical vertex function'', ${\cal V}^{(3)}_\Delta$, given by a sum
\begin{equation}\label{eq:loopV3}
{\cal V}^{(3)}_\Delta = \sum_{(X,Y,Z)_\Delta} \Theta_X \Theta_Y \Theta_Z \, G^{(3)}(X,Y,Z),
\end{equation}
over \emph{all} possible triangles $(X,Y,Z)_{\Delta}$ that correspond to (i.e. are dual to) $\Delta$.

In any cone $C$, only a single term in ${\cal V}^{(3)}_\Delta$ survives, and gives $G^{(3)}(X,Y,Z)$ for the one triangle in $C$, $(X,Y,Z)_\Delta$, that is dual to $\Delta$. Taking the product over all $\Delta$ then weights each cone $C$ with the correct numerator factor associated with the cubic Langrian ${\cal L}^{(3)} = G^{(3)}$.

We can compare this formula for ${\cal \tilde N}^{(3)}$ with the ``product over all triangles'' formula for ${\cal N}^{(3)}$ in equation \eqref{eq:loopN3}. For simplicity, consider the two formulas at tree-level. ${\cal N}^{(3)}$ is the product of $O(n^3)$ factors, each of the form $(1 + g^{(3)} \Theta \Theta \Theta)$. In any given cone, only $n-2$ of these factors are non-trivial (i.e. not equal to $1$), and these give numerators for the Lagrangian ${\cal L}^{(3)} = 1 + g^{(3)}$. By contrast, the ${\cal \tilde N}^{(3)}$ defined above is a product of $(n-2)$ factors, ${\cal V}^{(3)}$, each of which is itself the sum of $O(n^2)$ terms. In any given cone, only one term in each ${\cal V}^{(3)}$ factors survives, again giving a product over $n-2$ factors. But this time the vertex factors compute numerators for the Lagrangian interaction ${\cal L}^{(3)}= G^{(3)}$, rather than ${\cal L}^{(3)} = 1 + g^{(3)}$. 

\section{Simple Tropical Numerators $\leftrightarrow$ Non-Polynomial ${\cal L}$}
\label{sec:simplenonpoly}

Moving beyond cubic Lagrangians, we now consider the amplitudes for scalar Lagrangians with arbitrary higher interactions, ${\cal L}^{(m)}$, for all $m$.\footnote{The case of Tr$(\Phi^4)$ theory at tree-level has some special simplifications, that leads to a rather different tropical description than the main thrust of our paper, which we describe in Appendix B.} For simplicity, we will again begin by finding tree-level formulas for the tropical numerator functions ${\cal N}$. These formulas easily generalize to higher loop order, as explained in the next section, below.

The most obvious way of extending our story for cubic interactions is to define tropical numerator functions ${\cal N}^{(m)}$ as products over $m$-gons in the surface. For example, consider
\begin{equation}
{\cal N}^{(4)} = \prod_{(a,b,c,d)} (1 + g^{(4)}_{a,b,c,d}(X) \Theta_{X_{a,b}} \Theta_{X_{b,c}} \Theta_{X_{c,d}} \Theta_{X_{d,a}}  (\Theta_{X_{a,c}} X_{a,c} + \Theta_{X_{b,d}} X_{b,d}) ),
\end{equation}
where the product is now over all quadrilaterals, $(a,b,c,d)$. Once again, we can see that ${\cal N}^{(4)}$ factorizes correctly and therefore defines amplitudes. In the limit $X \to 0$, for some $X$, the factor for any quadrilateral $(a,b,c,d)$ that crosses $X$ goes to 1. Moreover, the factor for any quadrilateral that ``straddles'' $X$ (i.e. has $X$ as an interior chord) also goes to 1. Note that the factor $(\Theta_{X_{a,c}} X_{a,c} + \Theta_{X_{b,d}} X_{b,d})$ is crucial for this. If we are sending $X_{a,c}\rightarrow 0$ in a cone with a quadrilateral $(a,b,c,d)$ that straddles $X_{a,c}$, then the factor $\Theta_{X_{a,b}} \Theta_{X_{b,c}} \Theta_{X_{c,d}} \Theta_{X_{d,a}}$ does not vanish in that cone. However, the factor $(\Theta_{X_{a,c}} X_{a,c} + \Theta_{X_{b,d}} X_{b,d})$ does vanish in the limit, since $X_{a,c}\rightarrow 0$ and $\Theta_{X_{b,d}} \rightarrow 0$ (because $X_{a,c}$ and $X_{b,d}$ are intersecting). 

\begin{figure}[t]
    \centering
    \includegraphics[width=0.5\textwidth]{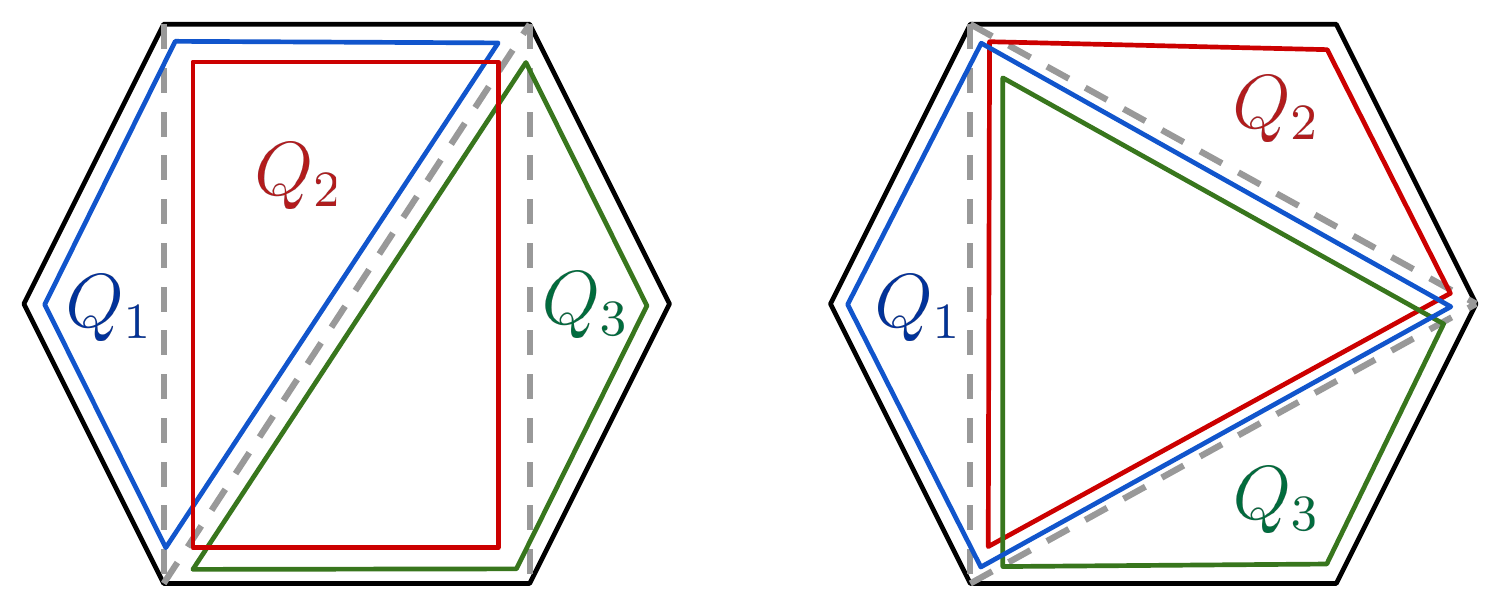}
    \caption{In any cone ${\cal N}^{(4)}$ detects the subset of quadrilaterals made of chords of the corresponding triangulation; this subset contains mutually overlapping ones.}
    \label{fig:overlap}
\end{figure}

But what theory does the function ${\cal N}^{(4)}$ correspond to? A first guess might be that it computes amplitudes for the quartic Lagrangian with ${\cal L}^{(3)} = 1$, and ${\cal L}^{(4)} = 2 g^{(4)}$. Indeed, the factors $(\Theta_{X_{a,c}} X_{a,c} + \Theta_{X_{b,d}} X_{b,d})$ can reproduce the quartic Feynman rule for this theory by killing the interior propagators of the quadrilateral $(a,b,c,d)$. But here we encounter an important difference from the case of cubic interactions. As before, in any given cone $C$, the only nontrivial terms in ${\cal N}^{(4)}$ come from quadrilaterals made from chords of the triangulation $C$. But many of these quadrilaterals are mutually overlapping, as in figure \ref{fig:overlap}. By taking the product over all of them, we are therefore not reproducing what we get from a quartic Lagrangian.  The product defining ${\cal N}^{(4)}$ does not know how to paste together subsets of non-overlapping quadrilaterals into quadrangulations. This issue did not arise for the cubic case, because any complete set of non-overlapping triangles always defines a triangulation.

Nonetheless, the numerator function ${\cal N}^{(4)}$ factorizes, and so it \emph{does} define amplitudes. But it turns out that the corresponding theory always has a more interesting \emph{non-polynomial} Lagrangian. For example, consider the tree amplitudes produced by ${\cal N}^{(4)}$ when we set $g^{(4)}(X) = g$, for some constant $g$. To find the Lagrangian for these amplitudes, we compute the contact terms at every order. We set the cubic term, ${\cal L}^{(3)}$, to $1$. At higher points, consider restricting ${\cal N}^{(4)}$ to any cone, $C$. Every internal chord $X$ appearing in ${\cal N}^{(4)}$ belongs to a unique quadrilateral, and hence, using the $(X \Theta_X)$ factors, every cone $C$ contributes $g^{(m-3)}$ to the $m$-point contact term. At four points we get $2 g$, five points we get $5 g^2$, and so on. The $m$-point contact term is $C_{m-2} g^{m-3}$, where $C_{m-2}$ are the Catalan numbers. It follows that ${\cal N}^{(4)}$ computes the amplitudes for the Lagrangian
\begin{equation}
{\cal L}_{{\rm Catalan}} = {\rm Tr} \left[\Phi^3 + 2 g \Phi^4 + 5 g^2 \Phi^5 + 14 g^3 \Phi^6 + \cdots \right] = 
 {\rm Tr} \left[\Phi \frac{1 - 2 g \Phi - \sqrt{1 - 4 g \Phi}}{2 g^2} \right],
\end{equation}
which we call the ``Catalan Lagrangian''. 

An interesting family of theories is obtained by taking $g^{(4)}(1,2,3,4) = g(X_{1,3},X_{2,4})$ to be a function only of the internal chords of the quadrilateral. Again, for any cone/triangulation $C$, every chord of $C$, $X$, uniquely belongs to one quadrilateral formed by the (other) chords of $C$. So, for every $X$, there is also the unique ``mutation'' of $X$, $X^{{\rm flip}}$, corresponding to the other diagonal of the quadrilateral. It follows that ${\cal N}^{(4)}$ computes the amplitudes of the theory with interactions
\begin{equation}
    {\cal L}^{(m)} = \sum_{\text{m-gon triangulations } T} \left(\prod_{X \in T} g^{(4)}(X,X^{{\rm flip}})\right).
\end{equation}
For example, if we further specialize to the case where $g^{(4)}(1,2,3,4)= \frac{1}{2} (g(X_{1,3}) + g(X_{2,4}))$, for some function $g(X)$, the contact term from every cubic diagram is simply the product of $g(X)$ over all the $X$ in the triangulation. Summing over all the triangulations gives a function that is formally exactly the same as the tree-amplitude for the Tr$(\Phi^3)$ theory, but for the replacement of $1/X \to g(X)$: 
\begin{equation}
{\cal L}^{(m)} = {\cal A}^{(m), {\rm tree}}_{{\rm Tr}(\Phi^3)}\left[X \to \frac{1}{g(X)}\right].
\end{equation}
In the special case where $g(X)$ is a constant, the Tr$(\Phi^3)$ amplitudes just count the number of diagrams and we are back to the Catalan Lagrangian. It is amusing that the simplest ``quartic'' tropical numerator functions can produce the amplitudes for a theory whose Lagrangian is itself entirely determined by the amplitudes for the Tr($\Phi^3$) theory!

The ``Catalan Lagrangian'' theory we encountered illustrates a fascinating general phenomenon: these amplitudes can alternatively be obtained by shifting the kinematic invariants, as discussed in \cite{zeros}. Indeed, suppose we take the amplitudes for Tr$(\Phi^3)$ theory, and simply shift all the kinematic invariants by $\frac{1}{X} \to \frac{1}{X} + g$. The new amplitudes defined by this substitution clearly still factorize onto themselves on all the poles.  It is also obvious to see that the contact terms for the $m$-point amplitudes are the Catalan numbers. So the amplitudes for ${\cal L}_{{\rm Catalan}}$ can be given a different, simple tropical representation, merely by taking the standard tropical presentation for Tr$(\Phi^3)$ and performing this simple kinematic shift. In fact, the amplitudes for the Catalan Lagrangian with $g \to \frac{1}{2}(g(X_{1,3}) + g(X_{2,4}))$ can also be obtained in this way, by performing the shift $1/X \to 1/X + g(X)$!  This is reminiscent of the (much more interesting) way in which non-linear sigma amplitudes can be obtained from Tr$(\Phi^3)$ theory by an even simpler linear kinematic shift \cite{nlsm}, which also gives rise to a different tropical representation of its amplitudes \cite{tropnlsm}. Of course, in general the amplitudes for any arbitrary ${\cal L}$ or ${\cal N}$ cannot be obtained by simple kinematic shifts in this way.

Finally, note that we can likewise define factorizing numerator functions ${\cal N}^{(m)}$ for every $m\geq 3$, defined as products over all $m$-gons, by analogy with ${\cal N}^{(3)}$ and ${\cal N}^{(4)}$, for some arbitrary choice of functions $g^{(m)}$. For any such choice of functions, $g^{(m)}$, the product
\begin{equation}
{\cal N} = \prod_{m=3}^\infty {\cal N}^{(m)}
\end{equation}
is itself a valid, factorizing numerator function. Each such ${\cal N}$ computes the amplitudes of some scalar field theory. However, as we have seen above, the associated Lagrangians in general involve infinitely many higher-point interactions. We can think of the functions $g^{(m)}$ as defining an alternative basis for the space of all possible scalar theories. This basis is non-trivially related to the conventional description of the space of theories using Lagrangians with finite numbers of polynomial interactions.

\section{Defining $m$-gons for general surfaces}
\label{sec:mgons}

\begin{figure}[t]
    \centering
    \includegraphics[width=\textwidth]{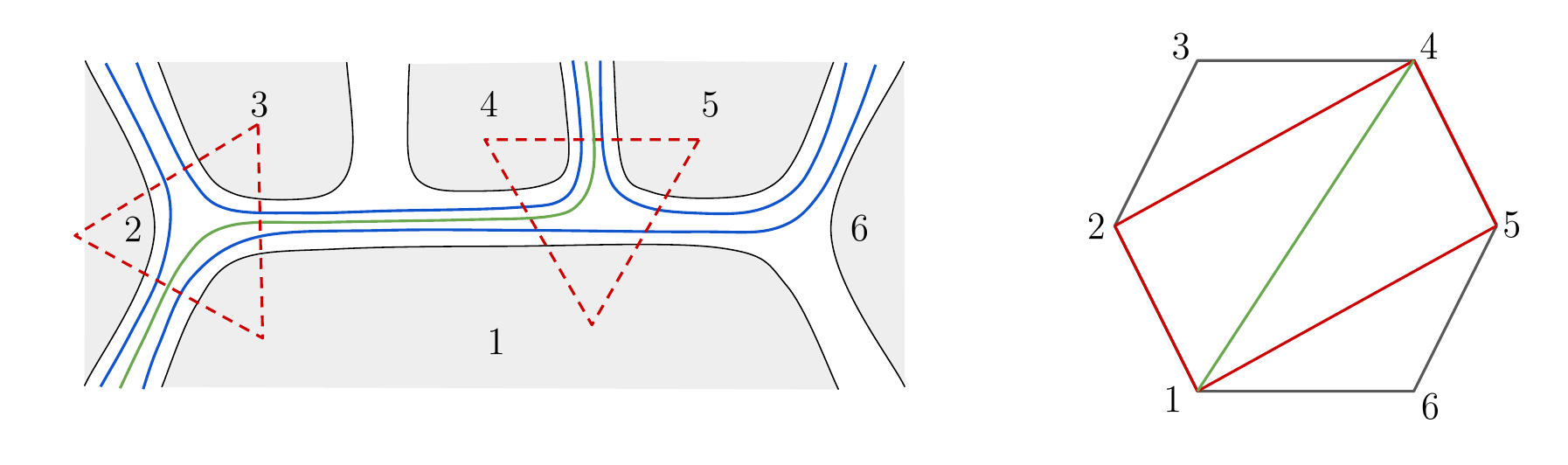}
    \caption{A quadrilateral defined by a cyclically ordered set of four curves $(X,Y,Z,W)$ that overlap on two vertices of the fat graph, as illustrated with an example here for $n=6$ at tree-level.}
    \label{fig:quadsurf}
\end{figure}

The previous section introduced, at tree level, numerator functions ${\cal N}$ for theories with higher valence vertices. To define these functions at all loop orders $L>0$, we have to first discuss how to define $m$-gons for general surfaces. The appropriate rule is a generalization of how triangles are defined for general surfaces, in section \ref{sec:cubic:triangles}. Once again, we fix some distinguished fatgraph $\Gamma$ (corresponding to a choice of base triangulation $T$ of the surface). 

To begin, consider quadrilaterals (i.e. $4$-gons). A quadrilateral is a cyclic 4-tuple of non-intersecting curves $(X,Y,Z,W)$ such that cutting $\Gamma$ along these curves gives a 4-point tree fatgraph. Just like the condition for triangles in section \ref{sec:cubic:triangles}, the condition that $(X,Y,Z,W)$ cut out a quadrilateral can be stated in terms of the base triangulation: the four curves must cut the corners of some {\it two} triangles $\Delta_1, \Delta_2$ in the base triangulation, $T$, and otherwise run parallel to each other. We write $(X,Y,Z,W)_{\Delta_1,\Delta_2}$ if $(X,Y,Z,W)$ is a quadrilateral dual to the two triangles $\Delta_1,\Delta_2$. Note that the two triangles, $\Delta_{1}$ and $\Delta_2$, dual to some quadrilateral, do \emph{not} need to share an edge. See figure \ref{fig:quadsurf} for an example at tree level.

The condition that $(X,Y,Z,W)$ is dual to $\Delta_1,\Delta_2$ can be expressed explicitly in terms of paths. Suppose that $(X,Y,Z)$ cut $\Delta_1$ and $(Y,Z,W)$ cut $\Delta_2$. If $\Delta_1$ is formed by roads $(i,j,k)$ and $\Delta_2$ from roads $(l,m,n)$, then $(X,Y,Z,W)$ have paths of the form
\begin{equation}\label{eq:XYZW}
X = A i L j \bar{B},~~ Y = B j L k C l L m \bar{D}, ~~ Z = E n L l \bar{C} k L i \bar{A},~~W = D m L n \bar{E},
\end{equation}
for some paths $A,B,C,D,E$. Just like for triangles, this equation can be reverse-engineered to \emph{generate} all the quadrilaterals on a surface, starting from pairs of triangles in the base triangulation.

Finally, note that if $(X,Y,Z,W)$ is a quadrilateral, there are precisely two curves that form the two diagonals of the quadrilateral. We will sometimes write $(X,Y,Z,W;S,T)$ for a quadrilateral with its two internal curves $S$ and $T$.

The above reasoning extends with trivial modifications to define $m$-gons on a general surface. An $m$-gon is an $m$-tuple of non-intersecting curves $(X_1,\ldots,X_m)$ that cut (i.e. are dual to) a set of $m-2$ triangles, $\Delta_i$, in the base triangulation.

\subsection{Simple Tropical Numerators Revisited}\label{sec:general:revisited}
Given the characterization of $m$-gons above, we can define simple tropical numerators ${\cal N}^{(m)}$ for any loop order $L$ as products over $m$-gons. For example, the class of numerators ${\cal N}^{(4)}$, defined in section \ref{sec:simplenonpoly}, become
\begin{equation}
{\cal N}^{(4)} =  \prod_{\Delta_1,\Delta_2} \prod_{(X,Y,Z,W;S,T)_{\Delta_1,\Delta_2}} (1 + g^{(4)}\, \Theta_{X} \Theta_{Y} \Theta_{Z} \Theta_{W}  (\Theta_{S} S + \Theta_{T} T)),
\end{equation}
where we now take the product over all quadrilaterals $(X,Y,Z,W;S,T)_{\Delta_1,\Delta_2}$ dual to all possible pairs of triangles, $\Delta_1,\Delta_2$, in the base triangulation $T$.

Alternatively, it is also possible to define numerator functions ${\cal \tilde N}^{(m)}$ as products over $m$-valent \emph{tropical vertex functions}, ${\cal V}^{(m)}$. These generalize the cubic tropical vertex functions introduced in section \ref{sec:cubic:vertices}. For example, we can define
\begin{equation}
{\cal \tilde N}^{(4)} = \prod_{\Delta_1, \Delta_2} {\cal V}^{(4)}_{\Delta_1, \Delta_2},
\end{equation}
as a product over all possible pairs of triangles $\Delta_1, \Delta_2$ in the base triangulation. Here, the cubic tropical vertex function is given by a sum
\begin{equation}
{\cal V}^{(4)}_{\Delta_1, \Delta_2} = \sum_{(X,Y,Z,W;S,T)_{\Delta_1, \Delta_2}} G^{(4)}(X,Y,Z,W;S,T) \Theta_X \Theta_Y \Theta_Z \Theta_W (S \Theta_S + T \Theta_T),
\end{equation}
over all possible quadrilaterals $(X,Y,Z,W;S,T)_{\Delta_1, \Delta_2}$ that are dual to a given choice of $\Delta_1,\Delta_2$. ${\cal \tilde N}^{(4)}$ computes the same amplitudes as ${\cal N}^{(4)}$, for the non-polynomial Lagrangians encountered above, if we make the identification $G^{(4)} \equiv (1 + g^{(4)})$. 

\section{Polyangulations from partitions}
\label{sec:partition}
The curve integral formalism naturally ``discovers'' all triangulations of a given surface, or, equivalently, all cubic graphs. The triangulations are efficiently enumerated using the tropical headlight functions, $\alpha_X$. It is natural to ask whether we can also enumerate all possible polyangulations in this way, and so discover all possible Feynman diagrams with higher-valence vertices? In this section, we explain how the curve integral formalism does naturally enumerate polyangulations. We find that every polyangulation is uniquely discovered from a partition of the set of $N=n-2+2L$ triangles, $\Delta_i$, in the base triangulation, $T$. This observation leads to new formulas for tropical numerators below, in Section \ref{sec:dictionary}.

As discussed in Section \ref{sec:mgons}, every $m$-gon is a tuple of curves $(X_1,\ldots,X_m)$ that are together dual to a set of $m-2$ triangles $\mathbf{\Delta} = \{\Delta_1,\ldots,\Delta_{m-2}\}$ in the base triangulation. Conversely, for any such set of $m-2$ distinct triangles, $\mathbf{\Delta}$, we can consider the set of all possible $m$-gons $(X_1,\ldots,X_m)_\mathbf{\Delta}$ that are dual to $\mathbf{\Delta}$. The $m$-gons in this set are pairwise incompatible: they all intersect each other. So, in a given cone/triangulation $C$, at most \emph{one} of the $m$-gons in this set is formed from the curves of $C$. In other words, for a fixed subset $\mathbf{\Delta}$ of $m-2$ triangles, $\mathbf{\Delta}$ defines, for each cone/triangulation $C$, either no $m$-gon or a \emph{unique} $m$-gon compatible with $C$.

To understand polyangulations, we need one further observation. Suppose we take two sets of triangles, $\mathbf{\Delta}$ and $\mathbf{\Delta}'$, having $m-2$ and $m'-2$ triangles, respectively. To $\mathbf{\Delta}$ we have the associated set of $m$-gons, $(X_i)_\mathbf{\Delta}$, and likewise to $\mathbf{\Delta}'$ we have the set of $m'$-gons, $(X_i')_{\mathbf{\Delta}}$. Then consider a cone/triangulation $C$ compatible with both an $m$-gon $(X_i)_\mathbf{\Delta}$ and an $m'$-gon $(X_i')_{\mathbf{\Delta}}$. The $m$-gon and the $m'$-gon intersect each other if and only if the two sets, $\mathbf{\Delta}$ and $\mathbf{\Delta}'$, have some triangles in common. In other words, pairs of compatible $m$-gons are uniquely dual to pairs of totally disjoint subsets of triangles.

It follows from this that polyangulations are uniquely dual to \emph{partitions} of the $N$ triangles $\Delta_i$ of $T$ into disjoint subsets, since a polyangulation is a maximal set of non-intersecting $m$-gons.\footnote{A polyangulation is \emph{maximal} if it comprises $m$-gons with valences $m_i$ such $\sum (m_i-2) = N$.} Conversely, fix a partitioning into subsets, $\{\mathbf{\Delta}\}$. Then, in any given cone $C$, this partition is either dual to a \emph{single} polyangulation, or it is dual to \emph{no} polyangulation (its image is ``empty''). By considering all possible partitions, in all possible cones $C$, we thereby discover all possible polyangulations of the surface!

Polyangulations are dual to Feynman diagrams with arbitrary valence vertices. So this result is a natural description of general Feynman diagrams in the curve integral formalism. We exploit this in Section \ref{sec:dictionary} to give formulas for tropical numerators. 

It is striking that all possible spacetime processes (i.e. all possible Feynman diagrams) emerge from such a simple combinatorial construction. The association between $m$-gons and subsets $\mathbf{\Delta}$ is canonical. Moreover, in order to generate polyangulations, we do not need to ``manually check'' how the curves of different $m$-gons intersect. Instead, it is enough to check that their corresponding subsets $\mathbf{\Delta}$ are disjoint. For the rest of this section, we illustrate the correspondence between partitions and polyangulations with a worked example.

\subsection{Example of the correspondence}
Let's illustrate the correspondence between polyangulations and paritions by studying the simple case of $n=6$ trees. As base triangulation, $T$, take the triangulation of the hexagon in figure \ref{fig:bijection}, with its four triangles $\{\Delta\} = \{\Delta_1, \Delta_2, \Delta_3, \Delta_4\}$.

Before discussing polyangulations, we begin by computing the $m$-gons associated to all subsets $\mathbf{\Delta}\subset \{\Delta_i\}$. For a general surface, we describe an $m$-gon as a cyclic tuple of curves. But, for this tree-level example, it is convenient to describe an $m$-gon simply by giving its vertices. For example, $(1,3,5,6)$ will denote the quadrilateral with these corners, bounded by the curves $(1,3), (3,5),(5,6),(6,1)$. 

First we list the triangles of the hexagon. Each of the 4 $\Delta$'s is dual to a set of possible triangles, and we find:
\begin{equation}
\begin{aligned}
    &\Delta_1 \to \{(1,2,3),(1,2,4),(1,2,5),(1,2,6)\}, \\
    &\Delta_2 \to \{(2,3,4),(2,3,5),(2,3,6),(1,3,4),(1,3,5),(1,3,6)\},  \\ 
    &\Delta_3 \to \{(1,4,5),(1,4,6), (2,4,5),(2,4,6),(3,4,5),(3,4,6)\},\\
    &\Delta_4 \to \{(1,5,6),(2,5,6),(3,5,6),(4,5,6)\} .
\end{aligned}
\end{equation}
We confirm here that all $20$ triangles of the hexagon are assigned to one (and only one!) of the $\Delta_i$. Now we list quadrilaterals, which we can discover as being dual to pairs of triangles $\Delta_i$. There are 6 possible pairs of triangles, and we find that they are dual to the following quadrilaterals: 
\begin{equation} 
\begin{aligned}
&\{\Delta_1,\Delta_2\} \to \{(1,2,3,4),(1,2,3,5),(1,2,3,6)\}, \\ 
&\{\Delta_1,\Delta_3\} \to \{(1,2,4,5), (1,2,4,6)\},\\ & \{\Delta_1, \Delta_4\} \to \{(1,2,5,6)\}, \\
&\{\Delta_2, \Delta_3\} \to \{(1,3,4,5),(2,3,4,5),(1,3,4,6),(2,3,4,6)\},\\ & \{\Delta_2, \Delta_4\} \to \{(1,3,5,6),(2,3,5,6)\}, \\ & \{\Delta_3,\Delta_4\} \to \{(1,4,5,6),(2,4,5,6),(3,4,5,6)\}.
\end{aligned}
\end{equation}
Once again we check that all $15$ quadrilaterals of the hexagon are dual to one (and only one) pair $\{\Delta_i,\Delta_j\}$. Now we list the pentagons, which are dual to sets of 3 triangles $\Delta_i$, and we find: 
\begin{equation}
\begin{aligned}
&\{\Delta_1, \Delta_2, \Delta_3\} \to \{(1,2,3,4,5),(1,2,3,4,6)\}, \\ & \{\Delta_1, \Delta_2, \Delta_4\} \to \{(1,2,3,5,6)\}, \\ & \{\Delta_1,\Delta_3,\Delta_4\} \to \{(1,2,4,5,6)\}, \\ & \{\Delta_2, \Delta_3, \Delta_4\} \to \{(1,3,4,5,6),(2,3,4,5,6)\},
\end{aligned}
\end{equation}
where we again check that all $6$ pentagons of the hexagon are uniquely assigned to one of the four triples of $\Delta$'s. Finally, we can also consider the set of all four $\Delta$'s, which is dual to the whole hexagon.

Note that a given subset $\mathbf{\Delta}$ does not necessarily have an image in every cone. For example, $\mathbf{\Delta} = \{\Delta_1, \Delta_4\}$ only ever corresponds to the quadrilateral $(1,2,5,6)$, and this quadrilateral is only compatible with those cones/triangulations that contain the chord $(2,5)$. 

Having enumerated all the $m$-gons, we can now consider the images of any partition of the 4 $\Delta$'s. Some examples are shown in figure \ref{fig:6part} (at the end of the paper). The images either correspond to complete polyangulations, or to partial polyangulations. In fact, some partitions never define a polyangulation in any cone. For instance, the pair of quadrilaterals dual to $\{\Delta_1,\Delta_3\}$ and $\{\Delta_2, \Delta_4\}$, respectively, always intersect. So this partition into two subsets has an empty image in every cone. However, discarding partial polyangulations, we also verify that every possible polyangulation arises as the image of one of these partitions. So, by taking the images of all possible partitions we discover every polyangulation of the hexagon.

\section{The Tropical/Lagrangian Dictionary} 
\label{sec:dictionary}

We have seen that simplest tropical numerators ${\cal N}^{(m)}$, studied in section \ref{sec:simplenonpoly}, compute amplitudes for \emph{non-polynomial} Lagrangians. In this section, we find tropical numerators ${\cal N}$ for \emph{any} Lagrangian, ${\cal L}$, with finitely many higher-point interactions. 

As we move around ${\bf t}$ space, the ${\cal N}^{(m)}$ defined in section \ref{sec:simplenonpoly} non-trivially compute all the allowed $m$-gon interactions that can occur in any cone. But these functions do not keep track of whether or not these $m-$gons are overlapping. To reproduce amplitudes for a polynomial Lagrangian, we need to pick out just those terms corresponding to non-overlapping $m$-gons. 

We will discuss two different ways of doing this. The first uses the familiar idea of ``Wick contractions''. We associate auxiliary variables $\Phi_X$ to every curve $X$, and explicitly ``glue'' edges of $m$-gons together by pairing up $\Phi_X$'s. The second method is more novel, and uses the correspondence (Section \ref{sec:partition}) between polyangulations and partitions of the triangles in the base triangulation. 

\subsection{General ${\cal L}$ at all orders: the Wick way}
To use Wick contractions to define numerators, introduce the auxiliary variables $\Phi_X$ associated with every propagator/curve, $X$, including the boundary curves. Let's begin with the case of quartic interactions. Define quartic vertex functions, $\Gamma^{(4)}(\Phi)$, that depend on the $\Phi$ variables, as a product over all quadrilaterals:
\begin{equation}
\Gamma^{(4)} =  \prod_{\Delta_{1},\Delta_2 \in T }\prod_{(X,Y,Z,W;S,T)_{\Delta_1,\Delta_2}} \biggl[ 1 + g^{(4)}\, \Phi_{X} \Phi_{Y} \Phi_{Z} \Phi_{W} \Theta_{X} \Theta_{Y} \Theta_{Z} \Theta_{W}  (\frac{1}{2} \Theta_{S} S + \frac{1}{2} \Theta_{T} T) \biggr],
 \nonumber
\end{equation}
for some function $g^{(4)}(X,Y,Z,W;S,T)$ defined for any quadrilateral $(X,Y,Z,W;S,T)$. Given this, we can define a quartic numerator function as 
\begin{equation}
{\cal N}^{(4)} = \prod_{X} \left(1 + \frac{1}{2}\frac{\partial^2}{\partial \Phi_{X}^2} \right) \left[\Phi_{1,2} \Phi_{2,3} \cdots \Phi_{n,1} \Gamma^{(4)}(\Phi) \right]_{\Phi \to 0},
\label{eq:Wick4}
\end{equation}
where $\Phi_{1,2} \cdots \Phi_{n,1}$ is the product over $\Phi_X$ variables for the boundary curves\footnote{This expression is for amplitudes with a single color-trace factor, corresponding to surfaces with just one boundary. For multi-trace factors, we would include a separate product of the $\Phi$'s associated with the boundary curves of each boundary component.}. We easily check that this new ${\cal N}^{(4)}$ factorizes correctly (analogous to the argument in section \ref{sec:simplenonpoly}). Moreover, the contractions of the derivatives with the $\Phi$-variables ensures that ${\cal N}^{(4)}$ is a sum of terms corresponding to \emph{non-overlapping} quadrilaterals. This is because the only terms that survive are those with exactly a factor $\Phi_X^2$ for each internal $X$ that contributes to that term. As a result, ${\cal N}^{(4)}$ computes the numerators for the quartic theory with ${\cal L}^{(4)} = 1 + g^{(4)}$.

The $m$-point couplings easily generalize from the $m=4$ case. To an $m$-field interaction Lagrangian, ${\cal L}^{(m)} = 1 + g^{(m)}$, we associate a vertex function $\Gamma^{(m)}(\Phi)$ defined as a product over $m$-gons, analogous to the definition of $\Gamma^{(4)}$. Then for some Lagrangian ${\cal L}$ with any (finite) number of higher interactions, we can compute its amplitudes using the tropical numerator function
\begin{equation}\label{eq:Ngeneral}
{\cal N} = \prod_{X} \left(1 + \frac{1}{2} \frac{\partial^2}{\partial \Phi_{X}^2} \right) \left[\Phi_{1,2} \Phi_{2,3} \cdots \Phi_{n,1} \prod_m \Gamma^{(m)}(\Phi) \right]_{\Phi \to 0},
\end{equation}
where we take a product over all $m$ for which the interaction ${\cal L}^{(m)}$ is nontrivial in the Lagrangian ${\cal L}$.

It is amusing that if we stop short of doing the Wick contraction, and consider the function in the square brackets in \eqref{eq:Ngeneral}, then we find a tropical function --- depending on the $\Phi$-variables --- that still factorizes correctly. These tropical functions therefore compute ``amplitudes'' that now depend on the variables $\Phi_X$. Indeed, we might be tempted to redefine  $g^{(4)} \Phi_{X} \Phi_{Y} \Phi_{Z} \Phi_{W} \to \hat{g}^{(4)}$. But this $\hat{g}^{(4)}$ now explicitly depends (via the $\Phi$-variables) on the quadrilateral, and so it cannot be interpreted as computing the amplitudes of a Lagrangian.

It is instructive to compare our form of the numerator ${\cal N}$  with the standard method of Wick contraction that directly generates the Feynman diagrams. Consider, for simplicity, the case of a quartic Lagrangian. The standard Wick contraction formula can be written (in momentum space) as
\begin{equation}\label{eq:WICK}
{\cal A}=\prod_{X} \left(1 + \frac{1}{2 X} \frac{\partial^2}{\partial \Phi_{X}^2} \right)  \left[\Phi_{1,2} \Phi_{2,3} \cdots \Phi_{n,1} \prod_{(X,Y,Z,W;S,T)} \left(1 + {\cal L}^{(4)} \Phi_{X} \Phi_{Y} \Phi_{Z} \Phi_{W} \right) \right]_{\Phi \to 0}.\nonumber
\end{equation}
There are several major difference between this formula and the tropical Wick formula, \eqref{eq:Wick4}. First, in the tropical Wick formula, the $\alpha_X$ functions break up the ${\bf t}$ space into piecewise linear domains that generate all the diagrams; the ``tropical Wick contraction'' only glues together a small subset of all the quartic couplings that are compatible with a fixed triangulation into the correct numerator factor for that triangulation. Hence, at large $n$, only $\sim n$ of the $(1 + \frac{1}{2} \partial^2/\partial \Phi^2)$ factors contribute to the formula at any point in ${\bf t}$ space. It is the integral over ${\bf t}$-space that then ``discovers'' the sum over diagrams. This is to be contrasted with the conventional Wick contraction, \eqref{eq:WICK}, which generates both the diagrams and the numerators at the same time. In this formula, \emph{all} $\sim n^2$ of the $(1 + \frac{1}{2 X} \partial^2/\partial \Phi^2)$ factors are relevant. Secondly, another important difference concerns loop integration. In the ``tropical Wick'' formula, we compute a tropical numerator function ${\cal N}$ that is defined over the whole of ${\bf t}$-space. While ${\cal N}$ may depend on the loop variables, the loop integration is still a Gaussian integral, and we can perform the integral ``once and for all'' for the whole amplitude. By contrast, in the conventional Wick formula, the loop integration must be done separately for each and every Feynman diagram.

As an aside, we briefly explain here how the results above can be extended to theories with multiple species $\Phi_I$; here, the $I$ index can be a stand-in for different flavors, spins, etc. The formula for ${\cal N}$, equation \eqref{eq:Ngeneral}, is largely the same in this case, but for a few minor changes. The couplings now carry $I$-indices, ${\cal L}^{(m)}_{I_1, \cdots, I_m}$. To define the $\Gamma^{(m)}$, we introduce new auxiliary variables, $\Phi_X^I$, that carry an $I$ index, and get contracted into ${\cal L}^{(m)}_{I_1, \cdots, I_m}$. Finally, to perform the Wick contraction, we need to make the replacement
\begin{equation}
    \left(1 + \frac{1}{2} \frac{\partial^2}{\partial \phi_X^2}\right) \to \left(1 + \frac{1}{2} P_{I,J} \frac{\partial^2}{\partial \phi_X^I \partial \phi_X^J}\right),
\end{equation}
where $P_{I,J}$ is the numerator of the propagator in species space.  

\begin{figure}[t]
    \centering
    \includegraphics[width=\textwidth]{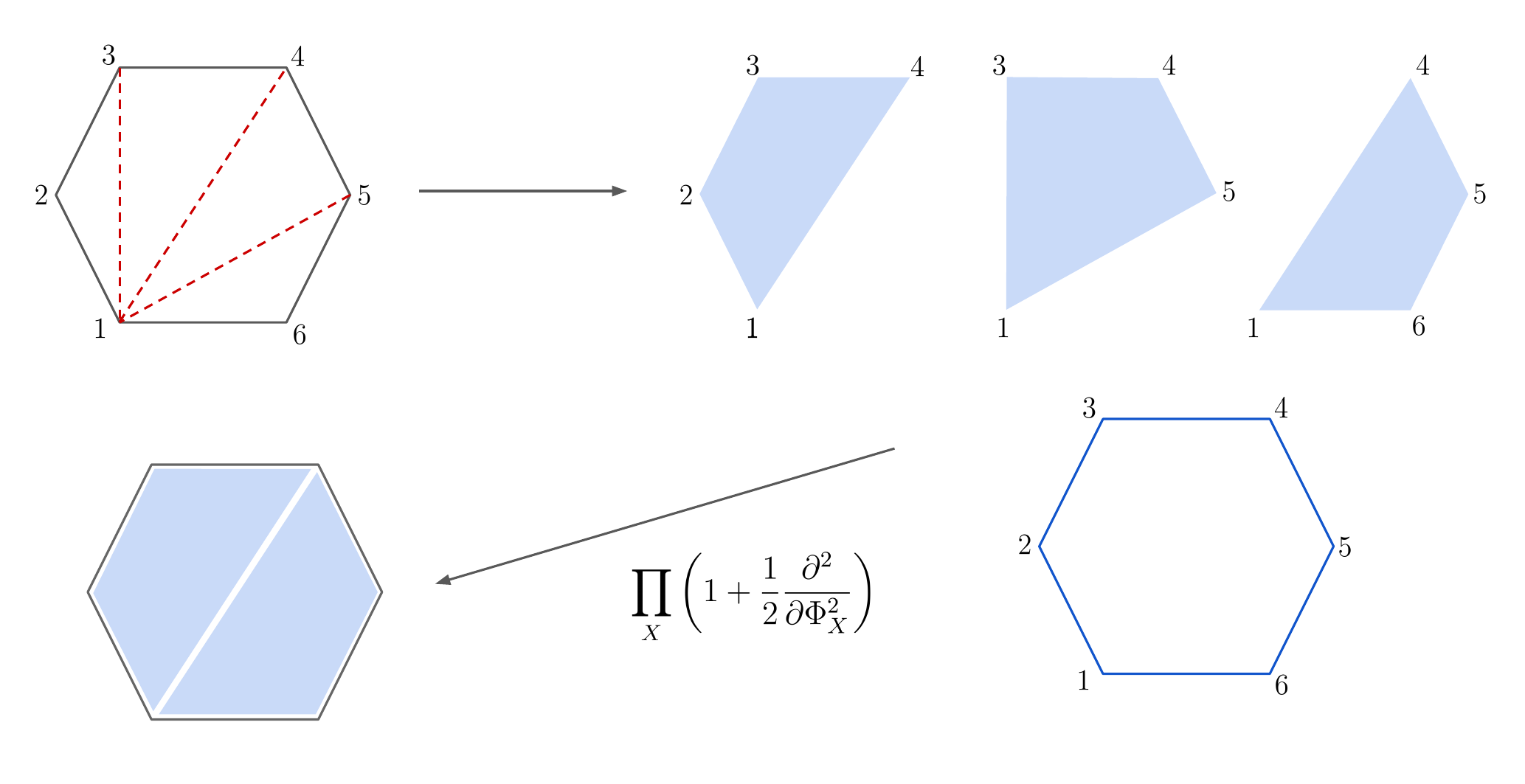}
    \caption{In any cone of the fan corresponding to a triangulation, $\Gamma^{(4)}$ produces all the quadrilaterals made curves of the triangulation, but with each edge $X_{i,j}$ decorated with $\Phi_{i,j}$. This is illustrated here for an example at $n=6$ tree level. We also have the product over all $\Phi_{1,2}\Phi_{2,3} \cdots \Phi_{n,1}$ denoted by the empty hexagon. The ``Wick contraction'' operator $\prod_{X} ( 1 + \frac{1}{2} \frac{\partial^2}{\partial \Phi_{X}^2})$ then acts on this product to identify edges and glue the internal and external polygons together, producing (in this case the unique) quadrangulation compatible with the underlying triangulation.}\label{fig:Wick}
\end{figure}

\subsection{General ${\cal L}$ at all orders: Tropical Vertices}
The tropical vertex functions, ${\cal V}^{(m)}_{\bf \Delta}$ defined in Section \ref{sec:general:revisited}, can also be used to define a numerator function ${\cal \tilde N}$ for any polynomial Lagrangian ${\cal L}$, but without using Wick contractions to manage the combinatorics. The idea is to leverage the correspondence between $m$-gons and sets of triangles, $\mathbf{\Delta} = \{\Delta_1,\ldots,\Delta_{m-2}\}$, in the base triangulation, $T$. This was studied in detail in Section \ref{sec:partition}. The key point was that two $m$-gons $(X_i)_{\bf \Delta}$ and $(X'_i)_{\bf \Delta'}$ can only be non-overlapping if all the two subsets ${\bf \Delta}$ and ${\bf \Delta}'$ are disjoint: i.e. have no triangles in common. We can therefore automatically produce products of compatible $m$-gons by simply summing ways to partition the $N = n-2+2L$ triangles of $T$.

To implement this, introduce $N$ variables $J_i$ --- one for every triangle $\Delta_i$ in $T$. Then, beginning again with the case of a quartic Lagrangian, ${\cal L}^{(4)}$, we can compute its amplitudes using the tropical numerator
\begin{equation}\label{eq:JN4}
{\cal \tilde N}^{(4)} =  \left. \prod_{i,j}\left[ 1 + {\cal V}^{(4)}_{\Delta_i,\Delta_j} J_i J_j \right]\right|_{J_1 J_2 \cdots J_N},
\end{equation}
where from the product we pull out the monomial $J_1 \cdots J_N$ in which every triangle $\Delta_i$ appears once. This guarantees that we only get those products of ${\cal V}^{(4)}_{\Delta_i,\Delta_j}$ corresponding to complete partition of the set of $\Delta$'s. In a given cone, $C$, these products turn into products over non-overlapping quadrilaterals compatible with the triangulation $C$.

Note that there is no contribution to ${\cal \tilde N}^{(4)}$ in $C$ from ${\cal V}^{(4)}_{\Delta_i,\Delta_j}$, whenever there are no quadrilaterals in $C$ dual to the triangles $\{\Delta_i, \Delta_j\}$.

Moreover, recalling the definition
\begin{equation}
{\cal V}^{(4)}_{\Delta_1, \Delta_2} = \sum_{(X,Y,Z,W;S,T)_{\Delta_1, \Delta_2}} G^{(4)}(X,Y,Z,W;S,T) \Theta_X \Theta_Y \Theta_Z \Theta_W (\frac{1}{2}S \Theta_S + \frac{1}{2}T \Theta_T),
\end{equation}
we see that ${\cal \tilde N}^{(4)}$ computes amplitudes for the quartic theory with ${\cal L}^{(4)} = G^{(4)}$.

Finally, for a polynomial Lagrangian ${\cal L}$ with any number of higher interactions, we can generalise \eqref{eq:JN4} to define a tropical numerator
\begin{equation}\label{eq:JNall}
{\cal \tilde N} =  \left. \prod_m \prod_{i_1, \cdots i_m} \left[ 1+ {\cal V}^{(m)}_{\Delta_{i_1}\cdots \Delta_{i_m}} J_{i_1} \cdots J_{i_m} \right] \right|_{J_1 J_2 \cdots J_N},
\end{equation}
where we take the product over all $m$ for which the interaction ${\cal L}^{(m)}$ is nontrivial in the Lagrangian ${\cal L}$.

It is interesting to contrast the tropical Wick formula for ${\cal N}$ with the tropical vertices formula for ${\cal \tilde N}$. The tropical Wick formula required us to consider the variables $\Phi_X$ associated to the edges of all possible $m$-gons, in order to then consistently glue the $m$-gons together into polyangulations. But the set of curves, $X$, for which $\Theta_X\neq 0$ varies as we move around in ${\bf t}$-space. So, in the tropical Wick formula, we have to recompute the ``Wick contractions'' as we move from cone to cone. The tropical vertices formula is more novel: extracting the coefficient of $J_1 J_2 \cdots J_N$ in \eqref{eq:JNall} is conceptually very different from ``gluing edges together''. Indeed, equation \eqref{eq:JNall} is simply generating all possible partitions of the set of triangles, $\{\Delta_1, \cdots, \Delta_N\}$. By contrast with the tropical Wick formula, this sum over partitions can be evaluated once and for all, and does not vary as we move around in ${\bf t}$-space. 
Another interesting difference between the formulas is that the form of the numerator ${\cal \tilde N}$ over ${\bf t}$-space explicitly depends on our choice of base triangulation, $T$. Whereas this is not true for the form of the numerator ${\cal N}$ given by the tropical Wick formula, which only depends on $T$ through the functions $\Theta_X$.

\section{Outlook} 
\label{sec:outlook}

We close by discussing some of the implications of the ideas in this paper and sketching some next steps.

We have seen a number of formulas for the numerator function ${\cal N}$ that are different as polynomials in $\Theta_X$'s, but which nonetheless agree as functions on ${\bf t}$ space. For instance the ``product over all triangles'' and ``tropical vertex'' formulae are very different as polynomials in $\Theta_X$, but they agree as functions if we identify $G^{(3)} = (1 + g^{(3)})$. In general, we have $\Theta_X^2 = \Theta_X$, and so in principle we can always work with functions that are at most linear in each $\Theta_X$. But certain monomials in the $\Theta$'s are vanishing, since $\Theta_X \Theta_Y = 0$ when $X,Y$ are crossing curves, and our different presentations of ${\cal N}$ differ by precisely such terms. In this vein, it would be interesting to understand the tropical avatar of the invariance of amplitudes under field redefinitions. Can such field redefinitions be naturally interpreted as total derivatives/variable changes in ${\bf t}$ space?

The constructions in this paper are adapted to work for generic theories, taking as basic ingredients the interaction vertices of Lagrangians. But one of the most dramatic facts about amplitudes exposed by the on-shell philosophy is that theories with very complicated looking Lagrangians, such as ${\cal N}=4$ SYM or gravity, often have stunningly simple amplitudes! The methods here are blind to this fact because, as given, the complexity of the Lagrangian is carried over into the tropical representations of the numerator functions. But there is an avenue for understanding more interesting tropical representations closer to the ideas we explored here. Recent work has shown that theories of special interest to the real world -- such as non-supersymmetric Yang-Mills theory and the non-linear sigma model describing pion scattering -- are related to the ``stringy'' Tr $\Phi^3$ theory in a more striking way, by certain simple linear shift of kinematical variables. We have already mentioned that this leads to a different tropical representation for the NLSM distinct from the approach of this paper. It is a pressing open question to find an analogous efficient tropical representation for Yang-Mills amplitudes, starting with the stringy representation in terms of ``scalar-scaffolded gluons'' given in \cite{gluons}. 

In this work we concentrated on field-theoretic amplitudes and their tropical representations. But all of our constructions lift naturally to stringy formulae as well. As shown in the appendix, the $\Theta_X$ variables have a nice interpretation as tropicalizing $(1 - u_X)$. Thus if we work with the stringy form of the Tr $(\Phi^3)$ amplitudes, and include our numerator functions with the replacement of $\Theta_X \to (1 - u_X)$, we obtain stringy generalization of the field theoretic amplitudes we are describing here. This strategy was sketched in \cite{killing}, where some simple classes of deformations of string amplitudes with consistent factorization on massless poles were described.

We have largely studied general Lagrangians for a single colored scalar $\Phi$. As observed in Section \ref{sec:dictionary}, the formulas can easily be extended to theories with multiple species of scalar, and it will be interesting to study the tropical numerators ${\cal N}$ for even more general theories. For instance, an importance case is to study the colored Yukawa theory with interaction Tr($\bar{\Psi} \Psi \Phi)$, which is very similar to the Tr$(\Phi^3)$ theory, but with the $\gamma^\mu P_\mu$ matrix structure for the numerator of fermion line propagators.

Finally, while we have focused our discussion on amplitudes for particles with color, it is natural to try and formulate seemingly even simpler theories, like that of a single real scalar $\sigma$, in a similar tropical way. Already the simplest case of the cubic $\sigma^3$ theory poses challenges to a tropical formulation, but there is reason for optimism: remarkably, all the diagrams for the $\sigma^3$ theory (as well as its interactions with Tr$(\Phi^3)$ via the coupling $\sigma$ Tr $(\Phi)$) are automatically present if we look closely enough at the $g$-vector fan of the surface, echoing the way in which closed strings are inevitably present in open string theory \cite{curveint}. The catch is that cones corresponding to the diagrams live are not top-dimensional, and related to this finding an effective way of modding out by the mapping class  group (which is relevant here even for tree-level processes!) is more challenging. It is interesting that a similar challenge arises if we attempt to describe higher-valence contact interactions in colored theories: our approach in this paper has been to ``blow up'' these vertices into cubic graphs, but we could alternately try to associate diagrams with higher-valence vertices with lower-dimensional cones in the fan. It would be interesting if a common strategy could be developed to attack both problems in a unified way. 

\acknowledgments
We thank Song He and Daniel Longenecker for useful discussions.
N.A.H. is supported by the DOE (Grant No.~DE-SC0009988), by the Simons Collaboration on Celestial Holography, and further support was made possible
by the Carl B. Feinberg cross-disciplinary program in innovation at the IAS. C.F. is supported by FCT/Portugal (Grant No.~2023.01221.BD).

\appendix 

\section{Lightning review of surfaces and headlight functions}
\label{sec:review}

In this section, we present a quick review of the basic facts about curves on surfaces that are necessary to compute the headlight functions $\alpha_X$ entering the curve integrals, as well as the \textbf{g}-vectors from which we compute the $\theta$ functions. We will provide a practical recipe on how to compute these objects. The interested reader can find further details into this formulation in the first section of \cite{curveint}, or look at section 1 of \cite{gluons} for a more detailed summary. 

While we can compute the vector $\textbf{g}_X$ using solely the information about the curve $X$, to compute $\alpha_X$ we need to first introduce the $u$ variable, $u_X$. $\alpha_X$ is then simply the tropicalization of $u_X$. 

\subsection{Step 1: Defining the surface}
\begin{figure}[t]
    \centering
    \includegraphics[width=\textwidth]{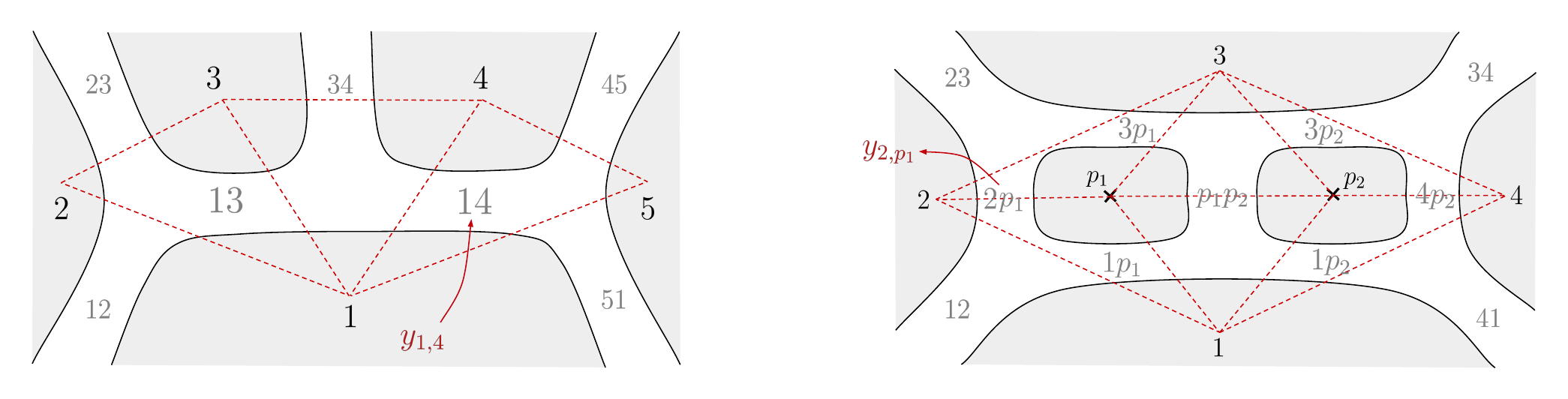}
    \caption{Two representative fat graphs, and the triangulations of the surfaces they define.}
    \label{fig:fatgraphs}
\end{figure}
The first step is to specify the surface, $\Sigma$, we are considering, $i.e.$ the order in the topological expansion. This is done by choosing one of the ribbon graphs contributing to the order in the topological expansion -- this we call the fatgraph $\Gamma$ \footnote{Providing a fatgraph, defines a triangulation of the surface, with chords $X$ corresponding to the propagators appearing in the fatgraph. Providing a triangulation is ultimately enough to specify the surface.}. In figure \ref{fig:fatgraphs}, we present two examples of fatgraphs one for the 5-point tree-level and the second one for the 4-point 2-loop.

In order to keep track of the curves on the surface/fatgraph, we label the edges of the fatgraph, $e$, and associate to internal ones positive variables $y_e$ (see figure \ref{fig:fatgraphs}). These will ultimately relate to the variables in which the curve integrals are defined via $t_e \coloneqq \log y_e$. Collectively, these edge variables $\textbf{t} \in \mathbb{R}^E$ give us the coordinates for the domain of integration of the curve integral associated with $\Sigma$.

\subsection{Step 2: Curves on the fatgraph, words and momenta}

The kinematic variables that the amplitudes depend on are in one-to-one correspondence to curves on the surface \footnote{This connection is explained in more detail in \cite{curveint,gluons}.}. Since we are using the fatgraph to describe the surface, the curves that we can draw correspond to those that enter via some boundary edge and exit elsewhere. For the case of the loop fatgraphs we can also have curves that spiral around the loops infinitely often, as we will see momentarily. 

Next, we need to find a way of encoding the information about a curve $X$ on the fatgraph. To do this we simply record its path on the fatgraph, by listing whether it turns left or right at each intersection. We do this by building a mountainscape where ``up''/``down'' stands for a ``left''/``right'' turn in a given intersection -- this is what we call the word associated with the curve, $W_X$ (see figure \ref{fig:WordFan} for the word of curve $X_{2,5}$ in 5-point tree-level (left)). 

There are two types of behavior the curves can have: they can either start (and end) on external edges of  $\Gamma$, or have an infinite spiral around the loops at their start/end, or both. For the first type, the respective words are finite while for the second one, due to the infinite spiral, we get infinite mountainscapes. 

Having learned how to define curves by words, we can use exactly the same data to attach a momentum $P_X^\mu$ to any curve $X$. To begin with, we assign momenta as we usually do to the internal edges of the fatgraph. 
Then, the momentum for any curve $X$ can be read off from its word $W_X$ by the rule
\begin{align*}
P^\mu_X = P^\mu_{\textrm{start}} + \sum_{\textrm{right turns}} P^\mu_{\textrm{incoming from left}},
\end{align*}
which requires summing over all occurrences of downward slopes/right turns in $W_X$, and adding the momentum entering from the left where $X$ is turning. In addition, we add the momentum of the external leg of $X$ from where the curve begins; if the curve begins with a spiral, we drop it from $W_X$ before applying the rule.

\subsection{Step 3: ${\bf g}$-vectors}

Already just with the information given in the words, $W_X$, we can read off the g-vectors, $\textbf{g}_X \in \mathbb{Z}^E$ (that uniquely identifies the curve). A given word, written as a mountainscape, has a unique set of ``peaks'' and ``valleys'' -- which always correspond to internal edges of the fatgraph. Then we have
\begin{equation}
    {\bf g}_X = -\sum_{{\rm peaks }} {\bf g}_{{\rm peaks }} + \sum_{{\rm valleys}} {\bf g}_{{\rm valleys }}. 
    \label{eq:gvectors}
\end{equation}

Since the peaks and valleys always correspond to the internal edges, we have that the g-vector of a given curve, ${\bf g}_X$, can be written as a linear combination of the curves in the underlying triangulation/fatgraph according to the rule \eqref{eq:gvectors}.
 
\begin{figure}[t]
    \centering
    \includegraphics[width=\textwidth]{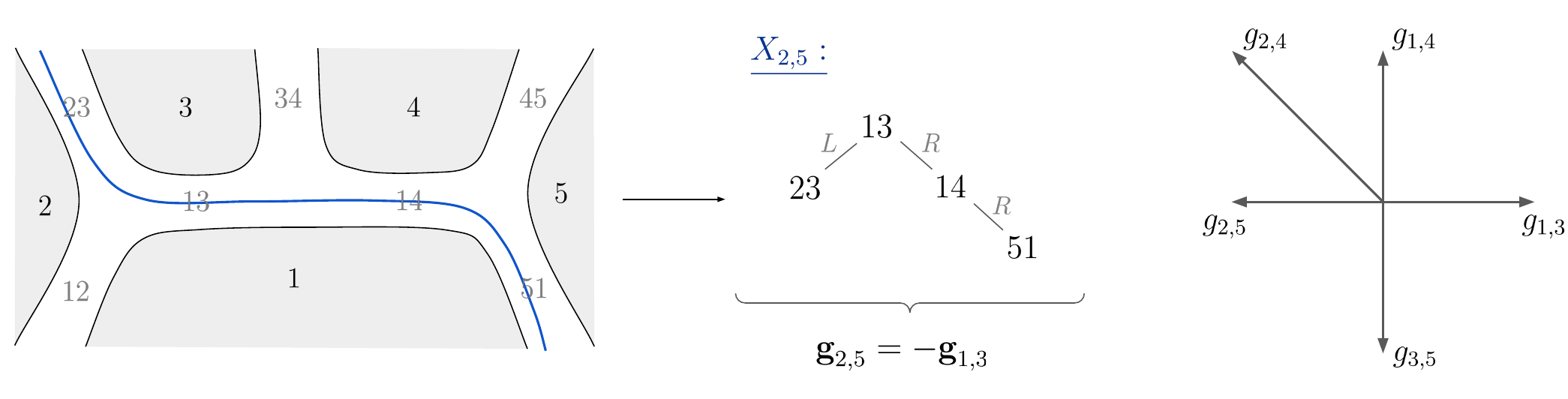}
    \caption{A curve on a 5-point tree fatgraph (left). The collection of g-vectors and cones for this fatgraph (right). Curves are labeled by twisting the endpoints counterclockwise so that the curve on the left corresponds to $X_{2,5}$ and has g-vector $g_{2,5}$ on the right.}
    \label{fig:WordFan}
\end{figure}

If we now consider a set of curves that defines a triangulation of the surface, $X\in C$, or, equivalently, that corresponds to the propagators present in a given diagram contributing to the order under consideration, then positive span of their g-vectors, $\{\textbf{g}_X\}_{X\in C}$, define a simplicial pointed cone in $\mathbb{R}^E$. A striking property of g-vectors is that the collection of all cones obtained this way form a \emph{fan}, i.e. they do not overlap with each other.
In figure \ref{fig:WordFan} (right) we show this phenomenon for the simplest case where the underlying surface is a disk with five marked points.
Already in this simple case, we can see that each cone is bounded by two g-vectors corresponding to curves that define triangulations of the disk with five marked points on the boundary. We have five different cones that then correspond to the five different triangulations. As we will see momentarily, the cones form the domains of linearity of the $\alpha_X$'s, appearing in the curve integral, which is the reason why when we integrate over the whole space it gives us the sum over all cubic graphs.

\subsection{Step 4: From words to $u$-variables}

As previously mentioned, we now want to define the $u$-variables associated with the curves $X$, from which we can extract the $\alpha_X$'s via tropicalization. 

To do this all we need is the words defining the curves. We start by defining a set of matrices associated to ``left''/``right'' turns from a given edge $y_i$:
\begin{equation}
	 M_{L}(y_{i}) = \begin{bmatrix}
		y_{i} & y_{i}  \\
		0 & 1 
	\end{bmatrix} \quad ;\quad M_{R}(y_{i}) = 
 \begin{bmatrix}
		y_{i} & 0  \\
		1 & 1 
	\end{bmatrix} .
\end{equation}
Then if for a given word, $W_X$, we can associate a 2$\times$2 matrix by multiplying these matrices according to the word, $M_X$. Note however that when we start in a boundary edge then the first turn does not have any $y_i$ associated with it, and so we evaluate the respective matrix at 1, $M_{L/R}(1)$. For the cases of spiraling curves, since the words are infinite, we get an infinite product of matrices. For now, let's consider a finite word and see what this procedure gives us for the curve $X_{2,5}$ depicted in figure \ref{fig:WordFan}:
\begin{equation}
    M_{X_{2,5}} = M_L(1) M_R(y_{1,3}) M_R(y_{1,4}) = \begin{bmatrix} 1+y_{1,3}(1+y_{1,4}) & 1 \\ 1+y_{1,3} & 1\end{bmatrix}.
\end{equation}

For a general curve, following the same procedure will produce a 2$\times 2$ matrix:
\begin{equation}
   M_X  =  M_{L/R}(1)M_{L/R}(y_1)M_{L/R}(y_2) \cdots  = \begin{bmatrix} M_{1,1}^X & M_{1,2}^X \\ M_{2,1}^X & M_{2,2}^X\end{bmatrix},
\end{equation}
where $M_{i,j}^X$ are polynomials in the $y_e$. Then the $u$-variable associated to curve $X$ is:
\begin{equation}
    u_X = \frac{M_{1,2}^X M_{2,1}^X}{M_{1,1}^X M_{2,2}^X}.
\end{equation}

So for example, for curve $X_{2,5}$, we get:
\begin{equation}
    u_{2,5} = \frac{1+y_{1,3}}{1+y_{1,3}(1+y_{1,4})}.
\end{equation}

\begin{figure}[t]
    \centering
    \includegraphics[width=\textwidth]{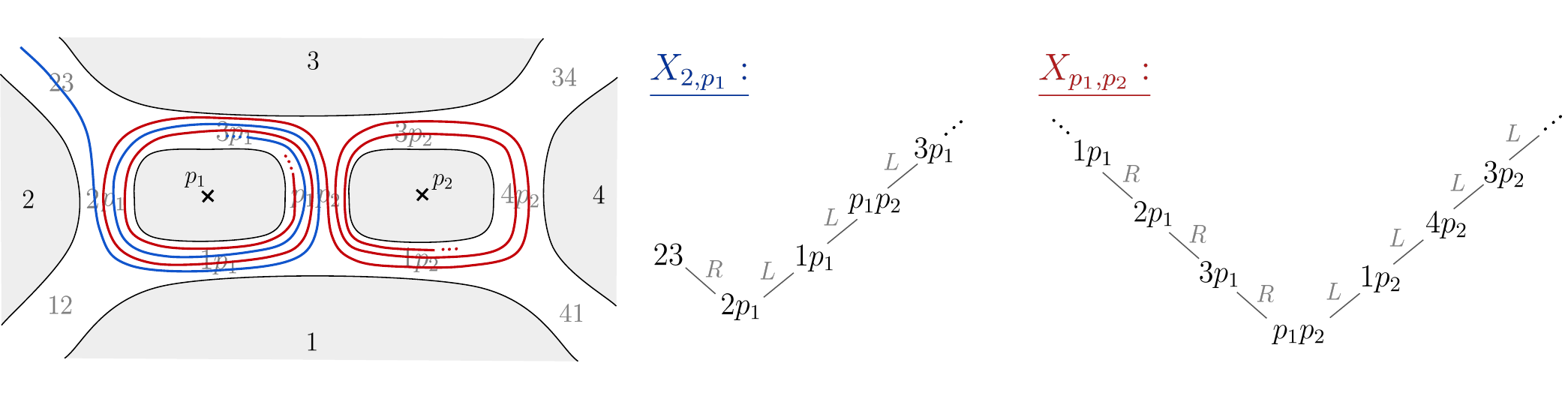}
    \caption{Two examples of spiraling curves. The blue curve spirals at the end, our convention is that end spirals are counterclockwise. The red curve starts and ends by spiraling; the spiraling, in the beginning, is clockwise.}
    \label{fig:Spirals}
\end{figure}

Let's now understand how we define $u$-variables for curves that spiral infinitely around the punctures. There are two possibilities: either the curve begins in a boundary edge and ends spiraling around some puncture, or begins and ends spiraling. In figure \ref{fig:Spirals}, we present two curves that illustrate both these possibilities, $X_{2,p_1}$ ends in a spiral, and $X_{p_1,p_2}$ that spirals both in the beginning and in the end. For both curves, the respective words are infinite (see figure \ref{fig:Spirals}, (right)), so we need to understand how to extract the $u$-variables in these cases. Let's start by considering the simpler case where there is only spiraling in the end. 

The defining feature of a word for such a curve is that it will end with the infinite repeating upward sequence:
\begin{equation}
    \begin{matrix}
     & &  & & &  &  & &  \udots \\
    & &  & & &  &  & y^{(p)}_1 &\\
    & &  & & &  &  \diagup & &\\
       & &  & & &  y^{(p)}_r & & &\\
       & &  & & \udots  & && &\\
       & & &y^{(p)}_2 & & & && &\\
       & &\diagup & & & && &\\
       & y^{(p)}_1& & & & & && &\\
       \udots  & & & & & & && &\\
    \end{matrix}
\end{equation}

So at the end of the word, we encounter the matrix $M_{{\rm spiral}} = M_L(y^{(p)}_1) M_L(y^{(p)}_2) \cdots M_L(y^{(p)}_r)$, which is raised to the $N$'th power with $N$ sent to infinity. Now it is easy to see that
\begin{equation}
M^N_{{\rm spiral}} = \left(\begin{array}{cc} Y_{{\rm spiral}}^N  & F_{{\rm spiral}} ( 1 + Y_{{\rm spiral}} + \cdots + Y_{{\rm spiral}}^{N-1}) \\ 0 & 1 \end{array} \right),
\end{equation}
where 
\begin{equation}
Y_{{\rm spiral}} = (y^{(p)}_1 \cdots y^{(p)}_r), \quad  F_{{\rm spiral}} = y^{(p)}_1 + y^{(p)}_1 y^{(p)}_2 + \cdots + y^{(p)}_1 \cdots y^{(p)}_r.
\end{equation}

Not that in order for the geometric series to converge, we must have $Y_{\text{spiral}}<1 \Leftrightarrow \prod_i y_i^{(p)}<1$. 

If we call the matrix product for the part of the word before we begin spiraling as
\begin{equation}
    M_{{\rm before}} = \left(\begin{array}{cc} a & b \\ c & d \end{array} \right),
\end{equation}
then if we take the $u$ variable for the matrix $M_{{\rm before}} M_{{\rm spiral}}^N$ and take $N \to \infty$ we find 
\begin{equation}
    u_{{\rm spiral}} = \frac{c (a F_{{\rm spiral}} + b (1 -  Y_{{\rm spiral}}))}{a (c F_{{\rm spiral}} + d (1 - Y_{{\rm spiral}}))}.
\end{equation}

We can also directly obtain this final expression for $u_{{\rm spiral}}$ without taking any limits, by defining a spiraling matrix as 
\begin{equation} 
M_{{\rm spiral \, end }} = \left(\begin{array}{cc} 1 & F_{{\rm spiral}} \\ 0 & 1 - Y_{{\rm spiral}} \end{array} \right).
\end{equation}

Then $u_{{\rm spiral}}$ is simply the $u$ variable associated with the matrix $M_{{\rm before}} M_{{\rm spiral \, end}}$. If we have a curve spiraling both at the beginning and the end, as in our example of $X_{p_1,p_2}$ connecting the two punctures at two loops, then we have a separate spiraling matrix at the beginning 
\begin{equation}
M_{{\rm spiral \, start}} = \left(\begin{array}{cc} 1 & 0 \\ F_{{\rm spiral}} & 1 - Y_{{\rm spiral}} \end{array} \right),
\end{equation}
and if the word sandwiched between the two spirals is associated with the matrix $M_{{\rm middle}}$ then the $u$ variables for the doubly spiraling curve is the one associated with the matrix $M_{{\rm spiral \,  start}} M_{{\rm middle}} M_{{\rm spiral \,  end}}$. 

\subsection{Step 5: From $u$-variables to $\alpha$'s and $\Theta$'s}
Now that we have given a recipe on how to obtain the $u$-variable for any curve on the surface, $u_X$, to obtain the corresponding headlight function, $\alpha_X$, we simply tropicalize $u_X$ which yields:
\begin{align}
    \alpha_X = - \mathrm{Trop}(u_X) =  \mathrm{Trop}(M^{C}_{1,1}) + \mathrm{Trop} (M^{C}_{2,2}) - \mathrm{Trop}(M^X_{1,2}) -\mathrm{Trop}(M^X_{2,1}).
\end{align}
``Tropicalizing'' means that we put $y_i = e^{-t_i}$, and examine the behavior of any polynomial in the $y_i$ as ${\bf t} \to \infty$. For instance consider $F = 1 + y_14 + y_14 y_13$. Then putting $y_{1,3} = e^{-t_{1,3}}, y_{1,4} = e^{-t_{1,4}}$ we examine the large $t$ behavior as  
\begin{equation}
F = 1 + e^{-t_{1,4}} + e^{-t_{1,4} - t_{1,3}} \to e^{\text{Trop}(F)},
\end{equation}
where
\begin{equation}
\text{Trop}(F) = {\rm Max}(0,-t_{1,4}, -t_{1,4} - t_{1,3}),
\end{equation}
simply records which monomial of $F$ dominates at large ${\bf t}$. The tropicalization of $F$ is a simplification of non-linear polynomials, capturing their asymptotic behavior by {\it piecewise linear} functions. 

So we see that the computation of $\alpha_X$ only requires the knowledge of the word $W_X$, and it is therefore entirely \emph{local} to the curve $X$, i.e. it does not depend on how $X$ fits with other curves in a particular triangulation. This is the basic ``miracle'' that turns the curve integral into a powerful representation of scattering amplitudes.

As an example, the headlight functions for the five curves on the fatgraph of figure \ref{fig:WordFan} are given by
\begin{equation}
\begin{aligned}
\alpha_{13}& =  t_{1,3} +  \max(0,-{t_{1,3}}),\\
\alpha_{14}& =  t_{1,4} - \max(0,-t_{1,3}) +  \max(0,-t_{1,4}, -t_{1,4} - t_{1,3}),\\
\alpha_{24}& = - \max(0,-t_{1,4}, -t_{1,4} - t_{1,3}) + \max(0,-t_{1,3}) + \max(0,-t_{1,4}),\\
\alpha_{25}& = - \max(0,-t_{1,4}) + \max(0,-t_{1,4},-t_{1,4} - t_{1,3}) \\ 
\alpha_{35}& = \max(0,-t_{1,4}).
\end{aligned}
\end{equation}
We can easily check that these correctly reproduce the fan of figure \ref{fig:WordFan}. We can also easily check the fundamental property that $\alpha_X({\bf g_Y}) = \delta_{X,Y}$. 

Additionally, we can use the matrix associated with the word, $W_X$ to directly compute the $\theta_X$'s. We start by noting that $\det W_X = \prod_{i \in W_X} y_i$, where the product is taken over all the $y_i$ appearing in $W_X$. So it follows from the definition of $u_X$ that
\begin{equation}
(1 - u_X) = \frac{\prod_{i \in W_X} y_i}{w_{1,2} w_{2,1}}.
\end{equation}
Now in any cone containing $X$, we have that $u_X \to 0$ asymptotically. So, in any cone containing $X$, $(1 - u_X)$ goes to 1. Hence, ${\rm Trop}(1- u_X) = 0$ in those cones that contain $X$, and it is nonzero otherwise. So we can define $\Theta_X$ 
\begin{equation}
\Theta_X = \begin{cases} 1 \, &{\rm if} \, \sum_{i \in W_X} (-t_i) - {\rm Trop}(w_{1,2}) - {\rm Trop}(w_{2,1}) = 0, \\ 0 \, &{\rm otherwise} \end{cases}.
\end{equation}

\section{Tr($\Phi^4$) trees}

Here we give a simple tropical description of the theory with quartic interaction $\frac{2 \lambda}{4}$ Tr($\Phi^4$) at tree-level; the factor of 2 is introduced in the coupling for convenience. Now Tr($\Phi^4$) trees have a special status in relation to a representation in terms of triangulations for a simple reason. The only propagators that can appear are for ``odd'' chords $X_{i,j}$ with $|i-j|$ odd. Consider a $2n$ point amplitude for simplicity at tree-level, the extension to general loops is trivial.  This has $(n-2)$ propagators. Any quadrangulation of the $2n$ gon with $(n-2)$ quadrilaterals can be completed to a triangulation, by filling in one of the two diagonals of each quadrilateral. Thus to get the amplitudes for the quartic theory with coupling $2 \lambda$, we simply take those triangulations that have precisely $(n-2)$ ``odd'' chords, take the product of the $1/X$'s for these chords, sum over all triangulations and multiply by $\lambda^{(n-1)}$. We can phrase this all using the $\Theta_X$'s trivially. We simply set all the $X_{i,j}$ with $|i-j|$ even to 1, and define the object 
\begin{equation}
N = \sum_{\substack{|i-j|\\ {\rm odd}}} \Theta_{X_{i,j}},
\end{equation}
to count the number of ``odd'' propagators in any cone. Then we have 
\begin{equation}
{\cal A}^{2n}_{\rm{Tr} \Phi^4} = \int_{S^{2n-4}} \frac{\langle t d^{2n-4} t \rangle \lambda^{n-1} \delta(N-n+2)}{\left(\sum_{\substack{|i-j| \\{\rm even}}} \alpha_{X_{i,j}} + \sum_{\substack{|i-j| \\{\rm odd}}} X_{i,j} \alpha_{X_{i,j}}\right)^{2n-3}}. \nonumber
\end{equation}

This expression correctly computes the Tr$(\Phi^4)$ amplitude, but in a sense, it is very inefficient. At large $n$, the vast majority of the triangulations don't have exactly $(n-2)$ chords with $|i-j|$ odd, and so have vanishing weight. Thus the tropical integrand has many ``holes'', and so has big fluctuations as we move around in ${\bf t}$ space. This is different than the case of Tr$(\Phi^3)$ theory where the cones all contribute and snugly fit together to cover the whole space.  While ``Stokes polytopes'' and their normal fans \cite{Banerjee:2018tun} can be associated with a subset of Tr($\Phi^4$) diagrams, it remains an open question whether there is some analog of the fan/polytope pictures for capturing the full Tr($\Phi^4$) amplitudes, in a way with far fewer ``holes'' than the expression given above.

 \bibliographystyle{JHEP}
  \bibliography{Refs}
\break 

\begin{figure}[t]
    \centering
    \includegraphics[width=\textwidth]{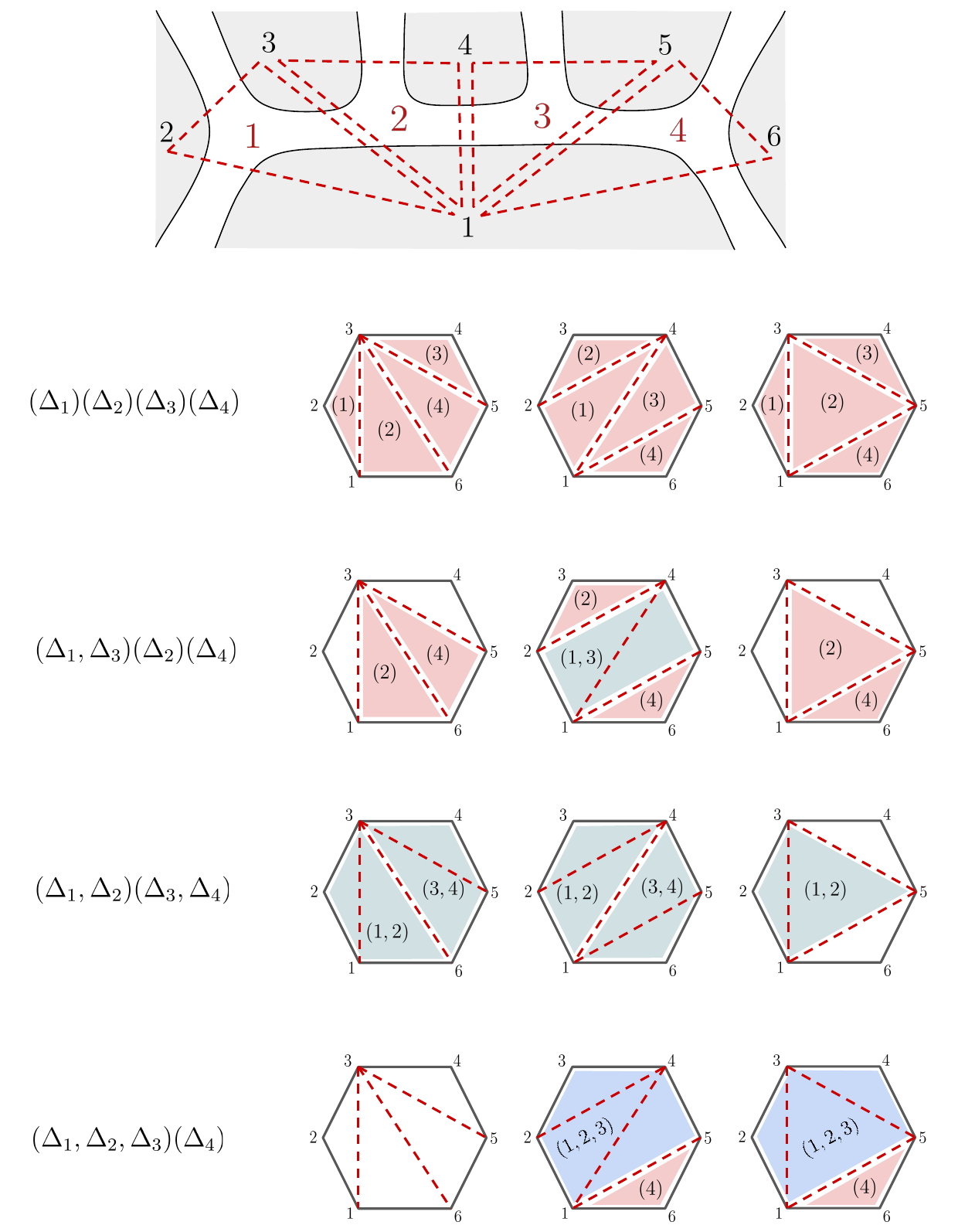}
    \caption{Images of a number of partitions of $(\Delta_1, \Delta_2, \Delta_3, \Delta_4)$ in three representative cones of the $n=6$ tree-level fan. In some cases, some or none of the groupings have images in the cones. But all images are non-overlapping, and whenever all subgroups have non-empty images, they give us a polyangulation made from curves of the cone. Ranging over all cones generates all possible polyangulations of the surface.}
    \label{fig:6part}
\end{figure}

\end{document}